\documentclass[conference]{IEEEtran}

\usepackage{lineno,hyperref}
\modulolinenumbers[5]
\usepackage[normalem]{ulem}
 \usepackage{color}
 \usepackage{graphicx}
\usepackage{epstopdf}
\usepackage{subfigure}
\usepackage{url}
\usepackage{amssymb,amsmath}
 
\providecommand{\eat}[1]{}

\begin{document}


\title{Storage Aware Routing for Generalized Delay Tolerant Networks}
 \author{\IEEEauthorblockN{Shweta Jain}
\IEEEauthorblockA{WINLAB\\
Rutgers University\\
North Brunswick, NJ 08902\\
Email: sjain@winlab.rutgers.edu}
\and
\IEEEauthorblockN{Snehapreethi Gopinath}
\IEEEauthorblockA{WINLAB\\
Rutgers University\\
North Brunswick, NJ 08902\\
Email: gopinath.sneha@gmail.com}
\and
\IEEEauthorblockN{Dipankar Raychaudhuri}
\IEEEauthorblockA{WINLAB\\
Rutgers University\\
North Brunswick, NJ 08902\\
Email: ray@winlab.rutgers.edu}}

\maketitle
\begin{abstract}
This paper presents a novel storage aware routing (STAR) protocol designed to provide a general networking solution over a broad range of wired and wireless usage scenarios. STAR enables routing policies which adapt seamlessly from a well-connected wired network to a disconnected wireless network. STAR uses a 2-Dimensional routing metric composed of a short and a long term route cost and storage availability on downstream routers to make store or forward routing decisions. Temporary in-network storage is preferred over forwarding along a path that is slower than average and opportunistic transmission is encouraged when faster than average routes become available. Results from ns2 based simulations show that STAR achieves $40-50\%$ higher throughput compared to OLSR in mobile vehicular and DTN scenarios and does $12-20\%$ better than OLSR in the static mesh case. Experimental evaluation of STAR on the ORBIT testbed validates the protocol implementation, and demonstrates significant performance improvements with 25\% higher peak throughput compared to OLSR in a wireless mesh network.
\end{abstract} 

%
\setcounter{page}{1}

 \section{Introduction} 	
In the United States, there were close to 280 million smart phone subscriptions in 2015 and the number is expected to cross 410 million by 2021~\cite{ericson_2015}. The consumer data demand is expected to be 9 exa-bytes per month in 2021 for smart-phone users alone. With several other applications over wireless links such as Internet of Things, it is expected that mobile and wireless devices connected to the Internet over wireless links will generate the majority of future Internet traffic load.  However, the current Internet protocol stack was not designed to handle  wireless connectivity and mobility. Industry and academic researchers have defined two major ways of dealing with this situation: innovations at the edge i.e., the access network or through a radical redesign of the core networking principles. Support for mobility at the edge is achieved through fast handoff, multi-homing, data pre-fetching, and exploiting access diversity such as the Google Project Fi~\cite{proj_fi}. The mobile network industry is working hard on new generation standards, 3G to 5G, that have progressively better support for mobility and handling of the demand for data. At the same time, a fairly large group of academic researchers have made progress in  clean slate re-design of the Internet to address native support for both mobility and security.  This Future Internet Architecture research initiative~\cite{ndn}~\cite{raychaudhuri2012mobilityfirst}~\cite{naylor2014xia} that started almost a decade ago is now near completion of the second phase.  MobilityFirst~\cite{raychaudhuri2012mobilityfirst}, Named Data Network~\cite{ndn}, eXpressive Internet Architecture~\cite{xia2011}, Nebula~\cite{nebula} and ChoiceNet~\cite{choicenet} are the five projects funded under the National Science Foundations FIA program~\cite{fia} in the United States. There are similar programs in Europe~\cite{eufia} and Asia~\cite{apan} as well. \eat{We contribute to the second initiative by designing a network protocol that adapts itself to diverse link conditions and mobility scenarios.

 The MobilityFirst architecture~\cite{raychaudhuri2012mobilityfirst}  puts mobility and wireless as the first-class citizens in their design. The Named Data Network~\cite{ndn} project reduces the transmission of redundant data through network routers through aggressive caching of named data within the network and delivering them to those who express ``interest'' in them. All architectures propose novel addressing and forwarding techniques~\cite{fara}~\cite{venkataramani2013design}~\cite{ndn} as baseline building blocks for architectural innovations. Radically new transport protocols were proposed such as the hop-by-hop transport~\cite{hop_umass}~\cite{cnf_arch}~\cite{Heimlicher} that finds its roots in the Cache and Forward Network architecture (CNF)~\cite{cnf_arch} and a named-data based transport in Content Centric Networking (CCN)~\cite{ccnx} which are predecessors of MobilityFirst and NDN respectively. Similarly addressing, naming and forwarding schemes were studied in depth in Forwarding directive, Association, Rendezvous architecture (FARA) which has been adapted in XIA to allow direct reachability to any principle such as content, host and service. These principles are free to move around in the network and yet be reachable through network layer mechanisms. 

  In the three architectures that we briefly describe above, named entities in the network are located using in-network lookup services. }
  
  The general approach in several FIA projects is to locate named entities such as content, service, host and other principles in the network using in-network lookup mechanisms to find the nearest point where the named entity can be found~\cite{vu2012dmap}. When combined with the mostly mobile Internet usage, one can expect a routing protocol that locates these entities must additionally build and maintain  various paths to nodes in the network with the  requirement that routes and choices of destinations are both optimized to support varying degree of disconnections. Here disconnection could be due to variable link quality observed in the wireless edge due to link conditions and host mobility  and may be extended to lookup delays, cache deletion, router outages as well as delays due to congestion. Even though routing is the main glue that can deliver  other architecture functions, this principle component of the network stack, in most FIA projects, is only loosely defined with the exception of MobilityFirst FIA.  Therefore, an important initial step is to design a single routing technique that can function well over a range of usage scenarios and provide seamless connectivity across a wide spectrum of stable to highly dynamic networks. From a protocol perspective, such a technique might as well be general enough to function under any network architectures, as long as some primary architectural requirements are satisfied. Both industry research in edge network enhancement and academic research in clean slate design can benefit from such a routing protocol. 
   
  In this paper, our goal is to  present a network routing protocol that can function well across a variety of mobile and wireless scenarios as well as deliver data under complete disconnections. We discuss, in Section~\ref{sec:related}, the generality of this protocol that can enable integration into various FIA architectures. We present comprehensive results from the design and analysis of this storage aware routing (STAR) protocol.  STAR uses link state advertisements so that network nodes can construct the connectivity graph. Each link in the graph data structure consists of the last known (short term) and historic (long term) packet transmission times and each node is marked with its storage availability. A breadth first search on this graph is used to compute multiple shortest paths to each destination. The path in which the sum of short term transmission times along each link is minimum is considered the best path. \eat{STAR then decides to transmit data along the best path if the short vs. long term transmission time is above the line $y = x-\beta$ and the minimum available storage along the path is more than $\gamma$. If the two path qualities fall under the line, data for the destination is stored and the decision is re-evaluated when the next path updates are received. }  STAR then decides to transmit the data along the best path if two conditions are satisfied. First: the short term transmission time along that path should be not much higher than the long term observations which is built as a moving average over an interval of time. Second, the storage availability on downstream routers must be above a threshold. If either condition is violated, STAR chooses to temporarily store the data instead of forwarding it along the path. In the scenario when the short term transmission time along the path is lower than the long term, STAR would opportunistically use that path to take advantage of the improvements, provided the storage criterion at downstream routers are met.  
  
 STAR was the basis of the the intra-domain routing scheme called Generalized Storage Aware Routing (GSTAR)~\cite{nelson2011gstar} in the MobilityFirst architecture. We note in Section~\ref{sec:related} that with some minor changes, it is also suitable as a single network protocol for Named Data Network instead of separate routing schemes that have been suggested to support ad-hoc~\cite{meisel2010ad}, VANET~\cite{grassi2013vehicular} and wired networks~\cite{hoque2013nlsr}.  We present the primary motivation that led to the design of STAR in Section~\ref{sec:motiv}. We then present the design of a storage router and choice of a transport layer in Section~\ref{sec:tp}. We describe the STAR protocol and present performance analysis in Section~\ref{sec:sim}. Related work, applicability to various FIA proposals and conclusions are presented in Sections~\ref{sec:related} and~\ref{sec:conclusion} respectively.

\section{Motivation}\label{sec:motiv}
The motivation behind STAR was to design a protocol that works in a range of wireless network types, ranging from well connected to disruption tolerant networks. It is well known that temporary path quality variations is an inherent characteristics of all wireless networks. These variations may be caused by fluctuations in signal-to-noise ratio (and hence the link speed), mobility, channel fading, as well as MAC layer congestion. In addition, energy constrained sensor and mobile phone networks may have intermittent disconnections due to low power and sleep mode operations. In general, cellular, Wi-Fi, multi-hop mesh, MANET, wireless sensor networks and tactical networks are all affected by some degree of intermittent disconnections. In this aspect, these networks display characteristics associated with DTN~\cite{Fall:2003} type disconnections although at shorter time scales ranging from a few seconds to minutes rather than several hours. In addition to disconnections, mobility and channel fading lead to temporarily poor channel conditions, causing the physical layer to drop the transmission data rates of active links. Mobile devices may have coverage from multiple networks (e.g. Wi-Fi, cellular and WiMax) with each handoff involving significant changes in path quality and end-to-end bandwidth. In contrast with complete disconnection, these phenomena have the effect of varying link bandwidth on the end-to-end path to a mobile device. Based on these considerations, we coin the phrase ``generalized Delay Tolerant Networks (gDTN)'' to describe the full range of wired and wireless network scenarios of interest. In this classification, DTN would correspond to a special case of challenged networks with long periods of disconnections. Similarly, well-provisioned wired networks represent the other extreme where channel quality is near-perfect and disconnections are rare.

\eat{FIA architectures with their name resolution services and storage capable routers also fit in this spectrum of generalized DTNs. As the cache entries in FIA routers expire leading to stale entries in the name resolution service, unexpected and variable delays are  introduced. Routers processing time is also variable as they now maintain other   data structures with variable size entries. For example, a single data packet arrival event in an NDN router may lead to multiple packet transmission  events to deliver data to each interface from which the interest for that packet was received. The forwarding information base (FIB) in routers in both NDN and MobilityFirst now contain multiple paths to each destination and hence, the processing time for selecting the  next hop node cannot be considered constant and negligible. A more extreme, albeit rare case is a router crash that potentially removes all the  reverse route information that NDN routing relies on for forwarding data back to the requester. Until the time the reverse routes are rebuilt from neighboring routers, the data in transit might as well be placed in temporary hold so that the work done to bring it at its current position is not totally futile. A corresponding situation in XIA is encountered when the router hosting an entity/principle goes offline and new sites needs to be discovered before proceeding. Such events are longer term disruption reminiscent of DTN like disconnections while the processing and congestion events demonstrate wireless mesh and ad-hoc network like characteristics.} 

Our goal is to design a unified network routing protocol that seamlessly connects the full set of gDTN scenarios including DTN and  wired networks.  At first, we will examine the characteristics and design considerations for routing protocols proposed for DTN as well as those for MANET and wireless mesh networks. We will then present the design of the unified approach.  The points to note are:
\begin{enumerate}
\item DTN routing is often based on historical observation of associations between nodes~\cite{Lindgren:03}. This choice is driven by the \eat{assumption} hypothesis that node mobility and connection/association probabilities between nodes can be predicted in the long term.
\item  MANET and mesh network routing protocols use first order metrics like hop based distance from the destination~\cite{aodv}~\cite{olsr-rfc}, transmission time~\cite{ett_metric}, data rate~\cite{edr_metric} and re-transmission count~\cite{etx}. The objective is to always find the best end-to-end path to the destination.
\item Since disconnection is a common phenomenon in DTN, routers in DTN store data locally while waiting for the destination route to become available. 
\item MANET and mesh network routing protocols react to temporary disconnections or route cost increase by starting route repair and alternate path search.
\end{enumerate}

We design STAR: a Storage Aware Routing protocol that borrows  principles from both DTN and MANET/mesh network protocols so that the combination is suitable for all gDTN scenarios. In this paper, we present a comprehensive design along with extensive simulation and emulation results that defined the STAR protocol and was later incorporated in the MobilityFirst GSTAR protocol~\cite{nelson2011gstar}. 
We make the following design choices for STAR:
\begin{enumerate}
\item{\bf{Storage Router:}}
We introduce the design of a storage router that allocates additional storage space for all layers of the protocol stack  (Section~\ref{sec:router}). 
\item{\bf{Two Dimensional Routing Metric:}} 
We design a two dimensional routing metric that makes routing decision based  on the current route cost and past history of the cost along the route (Section~\ref{sec:routing}). 
\item{\bf{Temporary Storage:}}
We borrow the DTN concept of temporary data storage when a destination route is unavailable and apply it to the general case where the routing cost is temporarily high as determined from the two dimensional routing metric (Section~\ref{sec:routing}). 
\end{enumerate}
We perform a detailed simulation based evaluation of STAR and present protocol validation and motivational experimental results from a proof-of-concept prototype implemented on the ORBIT~\cite{orbit05} testbed (Section~\ref{sec:sim}).
Results show that STAR performs well in mesh, MANET and DTN scenarios.  Although the storage and history features allow STAR to function in DTN scenarios, yet STAR is less comparable to various DTN routing protocols than it is to mesh network routing protocols. First, unlike most DTN routing protocols, STAR  relies on link speed and other physically observable qualities of the communication link can also be used while DTN routing protocols such as PROPHET~\cite{lindgren2012probabilistic} have used probability of contact as the main indicator of the link quality. Second, STAR does not use replication based techniques to improve the probability of data delivery to the destination. Most DTN routing protocols, from epidemic routing~\cite{vahdatand2011epidemic} to PROPHET~\cite{lindgren2012probabilistic}, rely on some amount of replication to improve  success. In sparse networks, given a suitable hypothesis regarding mobility and contact probability, any replication based routing is likely to outperform STAR which is a link state routing protocol. Conversely, random mobility scenarios can be built to make it hard to find a suitable hypothesis regarding contact probability. Since comparison between a link state routing protocol and replication and prediction based routing is rather subjective, in all our performance analysis, we refrain from comparing STAR with DTN routing protocols and instead use OLSR to benchmark the performance.
   
\section{Network Architecture}\label{sec:tp}
There are two important design choices for a gDTN network. First, we need a storage capable router and second, a transport protocol that is  best suited in our target scenario which consists of wireless and mobile devices. In this section, we present and justify our design choices. 
\begin{figure}[!t]
\centering
\includegraphics[width=0.5\textwidth]{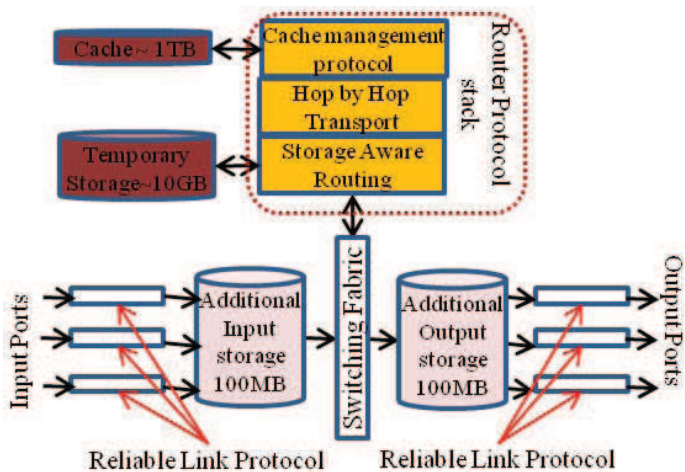}
\caption{Storage Router}
\label{cnf_router}
\end{figure}
\subsection{Storage Router}\label{sec:router}
Given the rapidly decreasing cost of semiconductor memory and hard drives, it is economically viable to build routers with large storage capacities. This extra storage can be used for in-network caching as well as for storing large content in transit. Figure~\ref{cnf_router} presents an example design of such a storage router. This router has a terabyte capacity cache to support in-network caching of content, 1GB space for network layer storage and 100s of megabytes for additional link layer input and output buffers. Such storage routers have also been suggested in the Cache and Forward~\cite{cnf_arch} architecture to support in-network caching and temporary storage when the end user disconnects during a content download process~\cite{PIMRC}.  In practice, such a router can also be built using the open-flow controller~\cite{openflow} with attached disk or solid state storage and the routing protocol may be applied to the open-flow controller overlay network. This requirement for a storage router is not too far fetched. Most FIA architectures are assuming such storage-routers to become more common than the ones that simply find routes, store forwarding information and push packets around in the networks. Both MobilityFirst and Named Data Network routers have significantly large storage assumptions to cache content and network flow information. Our storage router  works well in both designs and may be used in either architecture. In this work, we will only use the network layer temporary storage space and the input and output storage spaces to present performance benefits due to the routing enhancements alone.
\subsection{Transport Layer}
 Prior research shows that TCP is inefficient in wireless networks in general~\cite{gerla_tcp_99} and performs poorly in mobile scenarios~\cite{vaidya_mobile_tcp}. Variations of TCP like I-TCP~\cite{ITCP} and CLAP~\cite{CLAP} have suggested that both downstream and upstream TCP connections between a wireless host and the wired network should terminate at the access point (AP) or a proxy gateway. The AP or proxy should then create a separate connection with the wireless host to facilitate the end-to-end communication. This idea seems proper for static wireless hosts but in mobile scenarios this technique would require the setup of a new proxy and Mobile IP based forwarding from the original proxy (home agent) to the new proxy (foreign agent)~\cite{perkins1997mobile}. Moreover, if there is a temporary disconnection, the TCP connection between the AP or proxy and the wireless host may timeout. We assume that there is no special mechanism at the end application and no in-network caching is used. Therefore, unless the mobile reconnects at the same AP or finds a route to the AP from the new point of connection, the data transfer process will need to restart from the original server. Clearly, TCP and its variations are not the best choices in mobile scenarios. Therefore, hop-by-hop based reliable and connectionless
transport protocol ideas have been designed and tested for wireless and mobile networks. There have been several proposals for the hop by hop transport protocol~\cite{cnf_arch}~\cite{hop_umass}~\cite{Heimlicher}  with the general idea to  provide transport layer reliability at every hop along the path to the destination. In these protocols, the transport layer data unit (TPDU) may be as large as 1GB and all network routers are responsible for reliably transferring the entire TPDU as a whole from one hop to another. The difference in various versions suggested in literature is mainly the reliability mechanism and whether the hop by hop reliability is implemented at a special link layer as in the Cache and Forward Architecture~\cite{cnf_arch} or the transport layer implements batch acknowledgement to account for packets in the TPDU~\cite{hop_umass}. This hop based transport naturally allows routers to cache the entire file, more efficiently compared to any end-to-end transport. In the latter case, the individual TPDUs will need to be specially marked to indicate caching and then filtered and aggregated at the caching router to re-construct the original file. 
\eat{
In the Cache and Forward~\cite{cnf_arch} transport protocol, the task of fragmentation of the large TPDU and reliable transmission of each fragment to the next hop is delegated to a reliable link protocol. The CNF link layer fragments the TPDU($\sim$1GB) into smaller batches ($\sim$100MB) and each batch is further broken down into maximum transmission units (MTU) that fit into the payload of a link layer frame. Each MTU is then transmitted reliably using the Selective Repeat ARQ scheme. To improve efficiency, instead of per frame acknowledgement, a batch acknowledgement (BACK) is sent after all frames in the batch have been transmitted. The BACK contains a bit vector to indicate which frames need to be retransmitted.  A transport layer acknowledgement is sent after the link layer at the next hop reassembles the TPDU and delivers it to the transport layer. 

 The HOP transport protocol~\cite{hop_umass} does not assume a reliability service from the lower layers. Therefore, fragmentation of the large TPDU and ARQ scheme to ensure reliable delivery of the fragments are implemented at the transport layer. However, the CNF and HOP transport protocol only differ in the layer at which fragmentation and per packet reliability are implemented. In principle both protocols primarily promote the same concept which is reliable hop-by-hop transport of large TPDUs from the source to the destination. }
 
\eat{The hop-by-hop concept is also a general idea on which the the NDN transport protocol is based. The NDN protocol essentially uses a hop-by-hop interest propagation through routers while the corresponding data packets follow the bread-crumbs dropped by the interest moving toward the data source. At every hop, the NDN transport  verifies data integrity and makes data caching decisions. In these respects, the NDN transport scheme, in principle,  is similar to the hop-by-hop transport although it uses shorter packets rather than large $\approx 1 GB$ chunks. NDN  makes better use of pipe-lining because routers forward packets immediately rather than aggregating them to reconstruct a large TPDU first. However, this comes at the cost of 100\% packet transmission overhead because for each packet in the file, there is a corresponding interest. Even though the interest packet might be much smaller than the data, this overhead can be a significant load on the router's memory. }

Performance of all hop by hop based transport models has been shown to be better than TCP in wireless and wired scenarios~\cite{hop_umass} as well as in wired networks with wireless endpoints~\cite{cnf_arch}. Some flavor of hop-by-hop transport is also present in the Named Data network (NDN). In NDN, interest packets are propagated hop-by-hop along routers and the corresponding data packets follow the reverse route setup by breadcrumbs left behind by the interest packets. NDN works well in wired networks~\cite{yuan2013experimental} and has been shown to have application in Vehicular Networks (VANET)~\cite{grassi2013vehicular} and ad-hoc networks~\cite{meisel2010ad}. Therefore, any hop by hop transport protocol or NDN transport protocol may be used with STAR. We chose a hop-by-hop transport protocol with batch acknowledgments for per-hop reliability in our simulation study as well as the testbed experiments. Adaptation of STAR as a routing protocol for NDN is left as future work. 
\section{STAR: Storage aware Routing}\label{sec:routing}
The Storage Aware Routing (STAR) protocol uses a two-dimensional routing cost metric that consists of both instantaneous (short term) and historical (long term) route costs to construct the best routes to the destination. It then considers the storage availability on downstream routers to make a store or forward decision. In this section, we will describe the protocol and algorithmic details, particularly (a) the route cost metric used in STAR, (b) the network topology discovery process, (c) the path computation algorithm and (d) the store and forward algorithm. 
\begin{figure}[ht]
  \centering
   \includegraphics[width=0.5\textwidth]{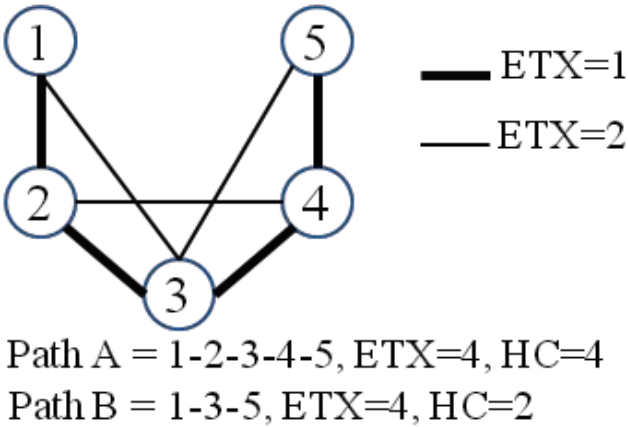}\\
  \caption{Fixed and variable routing costs on two alternate routes}\label{fig:etx_hops}
\end{figure}
\subsection{Routing Cost Metric}\label{subsec:ett}
In any multihop network, the cost of a route has two components: a variable component that primarily determines the quality of the path and a relatively fixed cost that depends on the length of the path from the source to the destination. Several quality metrics such as the transmission success probability (ETX~\cite{etx}), transmission rate in wireless networks (ETT~\cite{ett_metric} and EDR~\cite{edr_metric}) and queuing delay(improved ETT~\cite{ett_improved}) have been proposed in literature to measure spatio-temporal variations in link quality. Prior research, in wireless networks, show that routing protocols that minimize the variable cost such as ETX, ETT and EDR of the route provide better per flow throughput in comparison to when shortest path routing is used.  However, consider an example scenario shown in Figure~\ref{fig:etx_hops}, where two paths A and B of hopcount 4 and 2 respectively have the same variable cost ($ETX=4$). The probabilities that there will be 4, 3 and 2 transmissions when the shorter path (path B) is used are $1/4$ , $1/2$ and $1/4$  respectively($ETX=2 \Leftrightarrow P_{success}=1/2$ for each link). The probability of four transmissions along path A is always 1. Therefore, path A causes   as much interference and contention related overhead as path B with a probability $1/4$, 1.33$\times$ the interference compared to path $B$ with probability $1/2$ and 2$\times$ with probability $1/4$. Overall, there is a $3/4$ probability for path A to cause more transmissions and contention related overhead compared to path B. This simple example illustrates that there may be scenarios where selecting a minimum variable cost path may not be beneficial for the network even though a single flow might fare better. In addition, routes computed using variable costs as the metric tend to experience more churn due to frequent changes in the costs. Depending on the time-scale of the change, the resulting paths could exhibit a very high churn. Therefore, we use the hop distance as the primary link cost metric in STAR. However, hop distance fails to capture link state variations due to physical events such as fading, mobility, congestion and contention that lead to fluctuations in link quality. Therefore, we use hop distance to compute multiple paths to each destination and then use a second metric to allow STAR to rank paths on the basis of such variations. We avoid paths that are much longer than the shortest length path. Therefore, the longest alternative path in our work is bounded to one hop more than the shortest path. 

 We propose an expected packet transmission time (EPTT) as a variable link cost metric. We define EPTT in terms of the relation between the Signal to Interference and Noise Ratio (SINR) and the Bit Error Rate (BER) of the received
signal~\cite{rappaport-book}. When the SINR between a transmitter and receiver is high, transmitters may use high bit rate
modulation. Wireless standards and the wireless card specification sheets define the best transmission rate given the value of SINR~\cite{link_adapt_ICC03} at the receiver. In
practice, this information may be made available from the network interface card. We use this information to calculate EPTT as the ratio of packet size in bits to the best transmission rate possible for the signal to be correctly received. Being a function of SINR, the EPTT metric accurately captures temporal variations in link qualities which may be due to channel fading, interfering transmissions as well as node mobility. \eat{Nodes use the EPTT information received from other nodes in the network to calculate a moving average of the EPTT along each route in order to record statistical variations with time. Both instantaneous and moving average of EPTT are used in the routing algorithm. The time-scale at which short term values of EPTT is recorded depends on the intervals at which heart-beat messages are sent by the routing protocol to probe for network connectivity. The long term EPTT is then in terms of integral multiples of this message interval. }

\subsection{Network Topology Discovery}
STAR uses the Optimized Link State Routing protocol (OLSR)~\cite{olsr} as its baseline. The network topology information is collected using local broadcast of ``Hello'' messages and network wide relay of ``Topology Control (TC)'' messages. In order to reduce the number of redundant broadcasts, only a subset of nodes called multi-point relays (MPR)~\cite{olsr} participate in  relaying TC messages. The message formats and MPR selection processes are documented in the OLSR RFC~\cite{olsr}. We essentially use the same messaging format as OLSR. In order to relay additional messages needed to compute the two dimensional routing metric, a ``Hello'' message sent by a node $n_k$ in STAR contains additional fields: (a) the one hop neighbor list and the best transmission rate $r_i$ each neighbor $n_i$ could use when communicating with $n_k$ and (b) the amount of available storage on node $n_k$. The transmission rate $r_i$ is calculated from the SINR of the last packet received from neighbor $n_i$. The available storage indicates the total storage capacity less the amount of data waiting to be forwarded. Similarly, the TC messages originated at node $n_k$ contains (a) its own available storage information and (b) information from the hello messages from all two-hop neighbors. Using the information learned from ``Hello'' and TC messages, all nodes construct their view of the network connectivity graph. They use the EPTT reported in the most recent messages as the short term EPTT along the link. Nodes also compute the long term EPTT of each link as a moving average of EPTT reported in subsequent ``Hello'' and ``TC'' messages. Each node also  includes its available storage information in all messages it originales. The information collected through these messages are used to populate a table that consists of tuples ``Destination $<$Hop distance, previous hop,short term cost, long term cost, storage at previous hop$>$'' for each known destination in the network.  The routing control messages are sent as broadcast and are never buffered. Therefore, each received message reflects the most current information available to the node. Older messages with stale information are never in circulation.   Furthermore, when three consecutive messages from node $i$  are not received by  node $j$ within the expected time duration, instead of deleting the link from its table, node $j$ sets the short term entry for the link $i-j$ to a very high value. The long term cost is frozen until the next update from node $j$ is received.  The scalability of STAR is comparable to that of any link state routing protocol as well as OLSR. Various extensions of OLSR have been proposed to improve the scalability by reducing the routing control traffic, such as by using hierarchical routing~\cite{ge2005hierarchical}, fish eye metric~\cite{nguyen2007scalability}, clustering~\cite{ros2007cluster}  and prediction~\cite{medina2010olsrp}. Any of these extensions are useful as the baseline control messaging protocol for STAR.

\subsection{Path Computation}
STAR searches the graph data structure built using the topology information gathered from control messages to find multiple paths to each destination in the network. The hop-count metric is use to compute  a vector of paths ${\rm I\!P_{i,j}}$ between each pair of nodes $(i,j)$. The multiple paths are upto $q$ hops longer than the shortest path. In our experiments, $q$ is set to 1. Since the goal is to find multiple paths, a modified breadth first search algorithm~\cite{Newman2001} is used to build a path vector ${\rm I\!P^n}$ instead of the standard Dijkstra's algorithm. Given a pair of source $S$ and destination $D$, the breadth-first search algorithm works as follows.
\begin{enumerate}
\item Assign node S distance 0 to indicate that S is 0 steps away from itself and assign $\infty$ as the distance to all other nodes. Set $d\leftarrow 0$, and initialize the tree $T\leftarrow S<0,S,0,0,\omega_S>$
\item For all $k\in N$, whose assigned distance is $d$, visit each vertex $i$ that is directly connected to $k$ but is not the parent of $k$ in the tree. If $i$ has not been assigned a distance, assign it $d+1$. Add $i\langle d+1,k,x_{ki},y_{ki},\omega_k\rangle$ as a child to node $k$ in the tree $T$. If $i$ already has a finite distance, it must exist on the tree as a child to another node. We again add the node $i\langle d+1,k,x_{ki},y_{ki},\omega_k\rangle$ as a child of node $k$ on the tree. Add the newly found path $S-i$ to the path vector ${\rm I\!P_{S,i}}$ if the difference between the minimum and maximum distances in the path vector ${\rm I\!P_{S,i}}$ is less than a threshold $q$.
\item Set $d\leftarrow d+1$.
\item If $D$ is assigned a distance and the difference between the minimum and maximum assigned distances in the path vector ${\rm I\!P_{S,D}}$ is less than $q$ then repeat from step 2 else stop. 
\end{enumerate}

If $l$ is the length of the largest alternate path from $S$ to $D$ and $m$  is the average node degree, then step 2 executes $N_l = \sum_{j=1}^{l}{m^j}$ times and puts the same number of nodes (with duplicates) in the trees. This tree, rooted at $S$, in the worst case has $D$ as the last leaf node. A reverse traversal from each leaf node to root $S$ builds $l$ paths, with cost $l$ each. Therefore, cost of a reverse traversal from each leaf  to $S$, after the tree is built, costs $l\cdot m^l$ and gives $N_l = \frac{m^l -1}{m-1}$ paths. Thus the cost to find paths to $D$ is amortized to $\frac{(N_l + l\cdot m^l)}{N_l} = O(ml)$.    
\subsection{Store and Forward Algorithm}
\begin{figure}[ht]
\centering
   \includegraphics[width=0.5\textwidth]{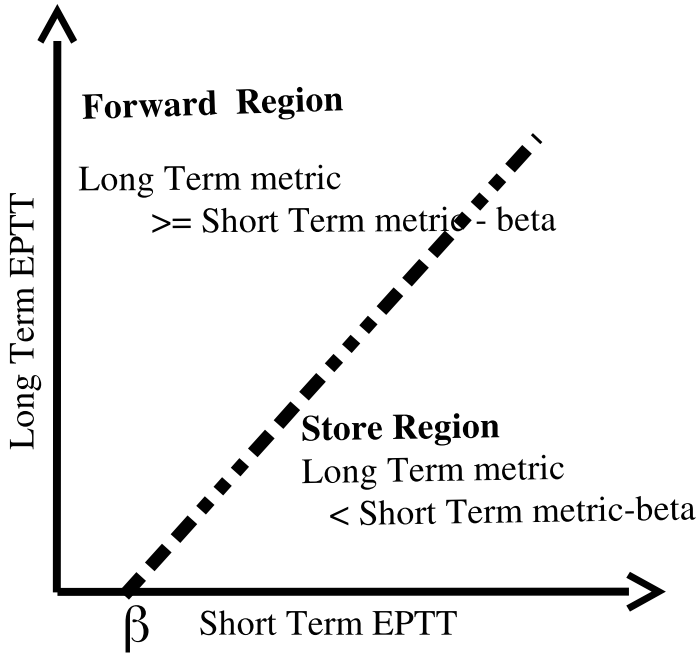}\\
  \caption{Forwarding Decision using the Two Dimensional routing cost metric}\label{fig:forwarding}
\end{figure}

All paths to the destination $D$ are sorted in increasing order of their short term EPTT. The long term EPTT ($y$) and short term EPTTs ($x$) of the best path i.e., the path with minimum short term EPTT are compared. We choose two thresholds, $\beta$ and $\gamma$ to determine the tolerance for paths that are slower than usual and for losses due to buffer overflow at downstream routers respectively. We describe a forwarding zone which is the region above the line $y=x-\beta$ and a storage zone that falls in the region below the line. A path that falls in the forward zone can be used to send files/data immediately provided that the downstream bottleneck router i.e., the one with the least storage available among all other downstream routers has sufficient space. The requirement is that the ratio of available to total space is atleast $\gamma$ and the space is sufficient to store the payload being transmitted. If a path is in the storage zone or if the downstream path has low space, file for the destination is sent to temporary storage on the current router.  Figure~\ref{fig:forwarding} illustrates this forwarding decision. 
  The choice of $\beta$  determines the tolerance for paths that are slower than usual. If $\beta$ is small, the system more aggressively avoids any path that is slower than than the long term EPTT. Therefore, systems where path quality fluctuates rapidly around a mean, a smaller $\beta$ is preferable. At the other end of the spectrum, if path quality fluctuates more slowly and hence may remain low for a long time, higher values of $\beta$ may be used to avoid delaying data transfer for too long. The choice of $\gamma$ controls how often files will be stored instead of being forwarded. This storage decision not only avoids the use of temporarily sub-optimal links but also provides a backpressure that reduces the rate at which traffic is injected into the network from the source. A higher value of $\gamma$ results in more aggressive flow control at the source while with lower values, the network pushes the data closer to the destination with the risk that it may be dropped enroute due to lack of space. For example, in heavily loaded networks, a larger value of $\gamma$ will lead to more effective flow control while under light load, the network will perform better with lower values of $\gamma$.  In STAR, the congestion information travels upstream at the same rate as periodic route updates travel and hence decision to choose alternate routes can be taken pre-emptively.  Since the back-pressure is in-built in routing, the  time scale of response to congestion is smaller compared to explicit congestion notifications (ECN)~\cite{rfc3168} in IP routers.

The forwarding decision in STAR only uses two indicators of path quality i.e., the expected transmission times and storage on routers. Other indicators of quality can also be used. For example, the GSTAR protocol in MobilityFirst factors in probabilities of various types of delays such as backpressure, channel contention and queuing to extend this forwarding decision~\cite{somani2012storage}.  Once the routes are computed, they are ranked based on the minimum delay metric before making the store and forward decisions.  Performance evaluations of GSTAR show some improvement over simple store and forward decisions in certain scenarios. 

\section{Performance Evaluation}
\label{sec:sim}
We evaluated the performance of STAR in a variety of gDTN scenarios using ns2~\cite{ns2} based simulations as well as on a proof-of-concept prototype on the ORBIT~\cite{orbit05} testbed. In this section, we will present the general architectural setup, define the performance metrics and present the performance results.
\begin{figure*}[ht]
\centering
\subfigure[HOP-STAR]{
\label{subCNF_L2_L3_L4}
\includegraphics[width=0.3\textwidth]{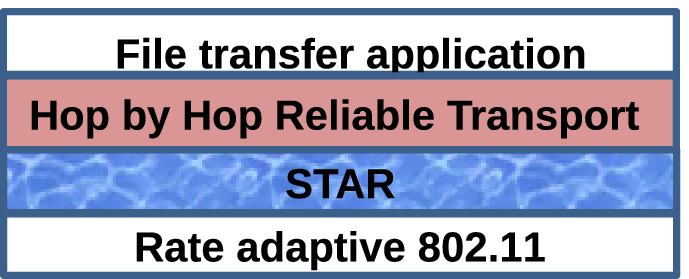}
}
 \subfigure[HOP-OLSR]{
\label{subCNF_L2_L4}
\includegraphics[width=0.3\textwidth]{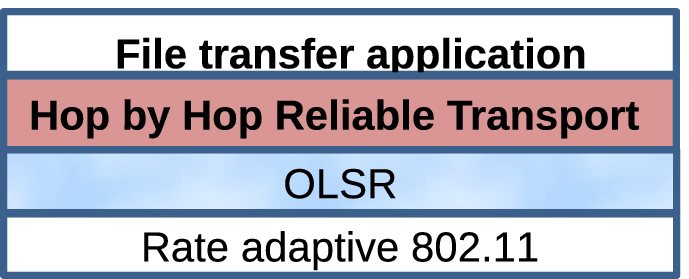}
}
\subfigure[TCP-OLSR]{
\label{subftp_stack}
\includegraphics[width=0.3\textwidth]{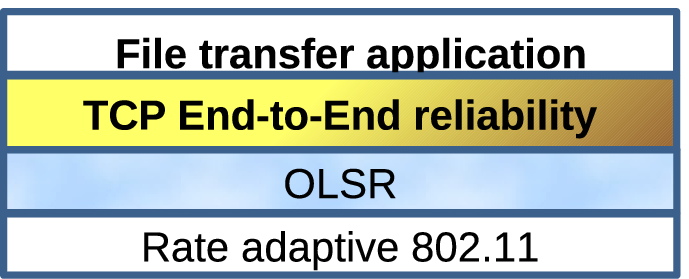}
}
\caption{Protocol stacks evaluated through ns2 simulations}
\label{expt_stack}
\end{figure*}
\subsection{Protocol Setup}

We have presented the protocol stack used in this work in Figure~\ref{expt_stack}. The stack shows the transport layer residing right above the network layer and below the application layer. In practice hop-by-hop transport protocols can be  implemented as a middle layer between application and the transport layer. In our implementation, the hop-by-hop transport provides a socket interface for the  application layer. The transport protocol fragments the application layer files (of size 1-2 GB) into transport protocol data units which are maximum 256KB in size. The transport layer then fragments the 256KB TPDU into smaller packets so that they can fit in link layer frames without undergoing IP fragmentation.  These smaller datagrams are sent through a UDP datagram socket and a single batch acknowledgment is requested from the next hop to confirm the receipt of the entire TPDU. Each fragment of the TPDU is encapsulated in network layer packets. The network layer performs link checking in accordance with the forwarding algorithm described in section~\ref{sec:routing} before sending the packets further down the stack. The next hop acknowledges by sending a bit-map indicating the sequence numbers of any packets that need to be retransmitted. When a zero bitmap is received, the segment has been successfully transmitted.  

\subsection{Performance Metrics}
 We use the following performance metrics for experimental evaluation in various static and mobile wireless scenarios.
\begin{enumerate}
\item \textbf{File streaming throughput}: In a DTN scenario, we calculate per hop throughput as the ratio of file size to the time taken for transmission across each link along the route. We then average the per hop throughput by adding throughput at each hop along the route and dividing by the number of hops. We report this average as the file streaming throughput. Delays incurred in queues and temporary storage are not used in this calculation.
\item \textbf{Average file delays}: We calculate the end-to-end delay for file transfers as a sum of total delay including the transmission time, queuing and storage delays incurred by files that are successfully delivered. We calculate an average end-to-end delay over all files that were delivered during the simulation.
\item \textbf{CDF of file delay}: We present the cumulative distribution of number of files delivered with respect to the end to end delivery delays for all files throughout the duration of each simulation run. 
\item \textbf{Average Network throughput}: The ratio of number of data bits transmitted through the network over the simulation duration is reported as the average network throughput.
\item \textbf{Four minute delay}: We present the fraction of files received with less than four minute end to end delay to show the peak number of files delivered at various offered loads. We chose this metric after observing the delay at saturation point which is around 240 seconds at 10Mbps offered load for all protocol stacks that were simulated. 
\end{enumerate}
\subsection{NS2 Simulations}
We used ns2 version 2.33~\cite{ns2} for all simulations.  We used an implementation of the hop-by-hop transport protocol with batch acknowledgement for hop-by-hop reliability (Liu et al)~\cite{sarnoff09}.  For comparison we used FTP data path at the application and TCP Tahoe implementations available in ns2. In OLSR experiments, we used the OLSR implementations by Ros et al~\cite{um-olsr}.  We experimented with three protocol stacks shown in Figure~\ref{expt_stack}. We will refer to these stacks as ``HOP-STAR'', ``HOP-OLSR'' and ``TCP-OLSR'' respectively. 

The HOP-STAR protocol stack in the simulator executes in the following manner. The hop-by-hop transport with per-hop batch acknowledgement runs on each router to ensure that the entire transport layer data unit (TPDU) is received reliably at every hop. Once a TPDU is successfully received at the next hop, it is queued in the temporary storage space available on the router. When the TPDU reaches the head of the storage queue and there is a suitable route available for the destination, it is dequeued from the storage buffer and encapsulated into a network layer datagram. The datagram is then placed in the router's output buffer, fragmented to fit into link layer frames and each fragment is transmitted to the next hop. After the entire TPDU has been transmitted, the sender  waits to receive an acknowledgement containing a  bitmap that indicates the frames that need to be re-transmitted. Once a bitmap with all bits set to 0 is received, the TPDU transmission to the next hop is considered successful and the next TPDU transmission may begin. 
   
\begin{table}
\centering
\caption{Simulation Parameters}
\begin{tabular}{|p{4cm}|c|}
\hline
\textbf{Parameter Name} &\textbf{Value} \\
\hline
Link layer frame size & 1024 bytes \\
\hline
Transmission data rates & 1, 2, 5.5, 6, 11 Mbps \\
\hline
Storage space & 2560 Kbytes \\
\hline
File Size & 256 Kbytes \\
\hline
Storage threshold $\gamma$ & 1280Kbytes \\
\hline
Moving average window size & 10 \\
\hline
Weightage on long term cost ($\beta$) & 0.9 \\
\hline
Simulation time & 1000s \\
\hline
Number of simulation runs & 10\\
\hline
\end{tabular}
\label{table:sim_params}
\end{table}
\subsubsection{Network, Traffic and Channel Model}
We evaluated our work in two stationary and two mobile scenarios. The simulation parameters are shown in Table~\ref{table:sim_params}.  The traffic model used in the simulations is as follows. Each node in the network periodically selects a random file and transmits it to a randomly chosen destination. The interval between subsequent file transfers generated by any node is exponentially distributed. We decrease the mean of the exponential distribution to increase the offered load in the network. We also experiment with a bursty traffic model in which files arrive at exponentially distributed intervals in bursts that last for 100 seconds. Consecutive bursts are followed by quiet periods of 200 seconds during which no new traffic is created at that node. The transmission bursts start and end randomly at different nodes so that active traffic is not synchronously generated on all nodes.

The 802.11b MAC and physical layer implemented by Chen et al~\cite{ns2_802_mac} was used in all simulations. This implementation provides a full featured implementation of wireless signal reception with cumulative signal strength from all transmissions that are in range to determine capture and collision. Thus, in all simulations, capture and collision are determined by the cumulative power of interfering signals that reach a receiver and not by pairwise comparison. For example, if there are $k$ simultaneous interfering transmissions during the reception of a frame that lasts from time $t$ to $t+\delta t$, the sum of signals from the $k$ interfering transmissions are considered when determining whether the received signal to interference and noise ratio (SINR) is high enough for successful reception of the frame. In addition, the interference power is monitored throughout the frame reception interval and any change in the interference level is factored into the computation of SINR. Thus, for successful reception of a frame, the SINR of the frame being received, computed cumulatively over all interferers, must stay above the receive threshold throughout the frame reception period.  This is more accurate compared to pair-wise SINR computation with each competing transmission which overestimates the packet reception probabilities. Preamble and header processing, packet capture and accurate model of the IEEE 802.11 MAC protocol have also been implemented. We added the auto rate adaptation~\cite{aarf} module to the 802.11 implementation and made necessary changes so that the SINR of received packets can be queried by the network layer to calculate the expected packet transmission time of the link. Since the transport layer implements per-hop reliability, 802.11 DCF using RTS and CTS messages is not used, however per-frame acknowledgement is still provided by the link layer.  We used the two-ray ground~\cite{Lee-book} based propagation model in all simulations with the exception of simulation 1 where we used an on-off propagation model in addition to the two-ray ground model. The transmission range in all simulations is set to 250m and carrier sensing range is 500m. Thus, in the absence of noise and interference, a receiver can be upto 250m away to receive a frame successfully. Under the same conditions, all other nodes that are within 500m of the transmitter can sense the carrier to determine that the channel is busy. All experiments were repeated for 10 runs and an average as well as confidence interval error bar are presented in simulation results. 
\begin{table}
\centering
\caption{Simulation 1(Linear topology): Results}
\begin{tabular}{|p{2cm}|c|c|c|c|c|c|}
\hline
\multicolumn{1}{|p{2cm}|}{\textbf{Parameters}} & \multicolumn{3}{|c|}{\textbf{OLSR}} &  \multicolumn{3}{|c|}{\textbf{STAR}} \\
\hline
Mean file arrival times (sec) & 50 & 10 & 1 & 50 & 10 & 1 \\
\hline
Average file delay (sec) & 0.678 & 1.047 & 2.936 & 1.634 & 10.883 & 5.518\\
\hline
Throughput (Mbps) & 0.302 & 0.548 & 1.417 & 0.337 & 0.897 & 1.765 \\
\hline
\end{tabular}
\label{table:sim1_results}
\end{table}
\begin{figure}[ht]
\centering
\includegraphics[width=0.5\textwidth]{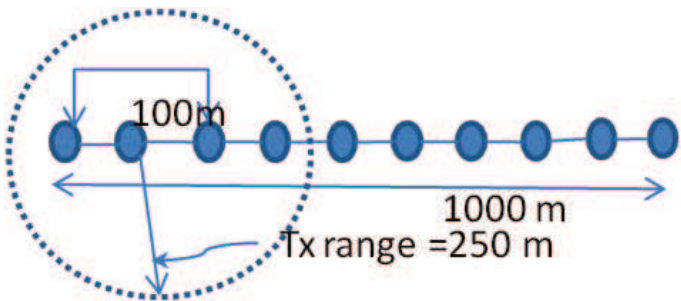}
\caption{Simulation 1 - Linear Topology with On-Off links:}\label{onoff_scenario}
\end{figure}
\paragraph{Simulation 1: Linear Topology with On-Off Links}
 We simulate a 10 node topology shown in Figure~\ref{onoff_scenario}. The goal is to show the benefits of using store-and-forward routing under varying link conditions. Therefore, in addition to the two-ray ground propagation loss model with capture and collision, additional variation in link conditions is simulated by switching the link rates from 1 to 11 Mbps and vice-versa at consecutive time intervals uniformly distributed between 20 to 50 seconds. The schedules and duration of these variations are randomly selected across different links. Results (see Table~\ref{table:sim1_results}) show that ``HOP-STAR'' achieves higher network throughput compared to ``HOP-OLSR'' but at the cost of high end-to-end file delays. The throughput gain is a direct result of reduction in contention for the wireless medium when nodes decide to store rather than transmit when the link rate is slower than the long term average. This indicates that the storage aware design is good for the network as a whole. At the same time, larger file delays were observed because files that incurred longer delays are eventually delivered successfully while in OLSR those files are likely to be dropped at intermediate queues due to transmission failures.
\begin{figure}[ht]
\centering
\includegraphics[width=0.3\textwidth]{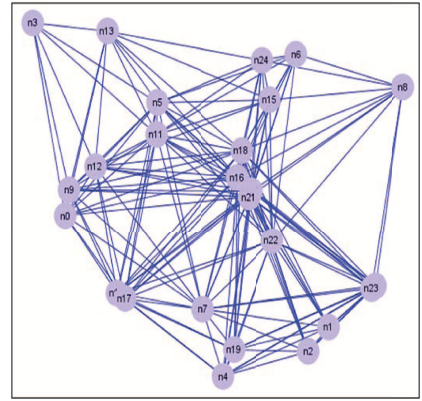}
\caption{Simulation 2 - Connectivity graph of nodes in the mesh network:}
\label{scenario}
\end{figure} 
\begin{figure*}[ht]
\centering
\subfigure[CDF of file delays: mean inter-arrival time of files is 5 seconds (offered load = 10Mbps)]{
\includegraphics[width=0.3\textwidth]{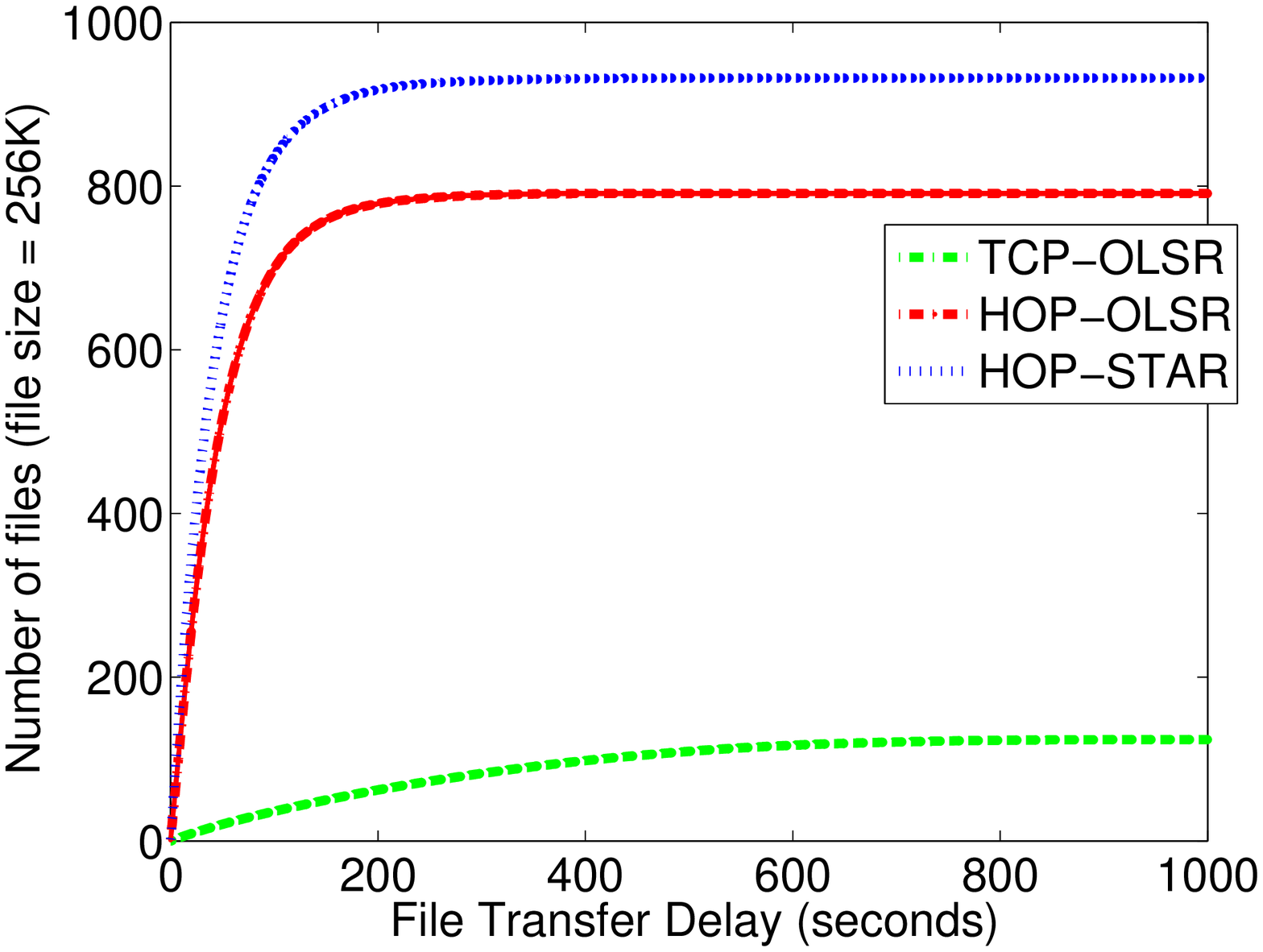}
\label{cdf}
}
\subfigure[Number of files delivered with less than 4 minutes delay]{
\includegraphics[width=0.3\textwidth]{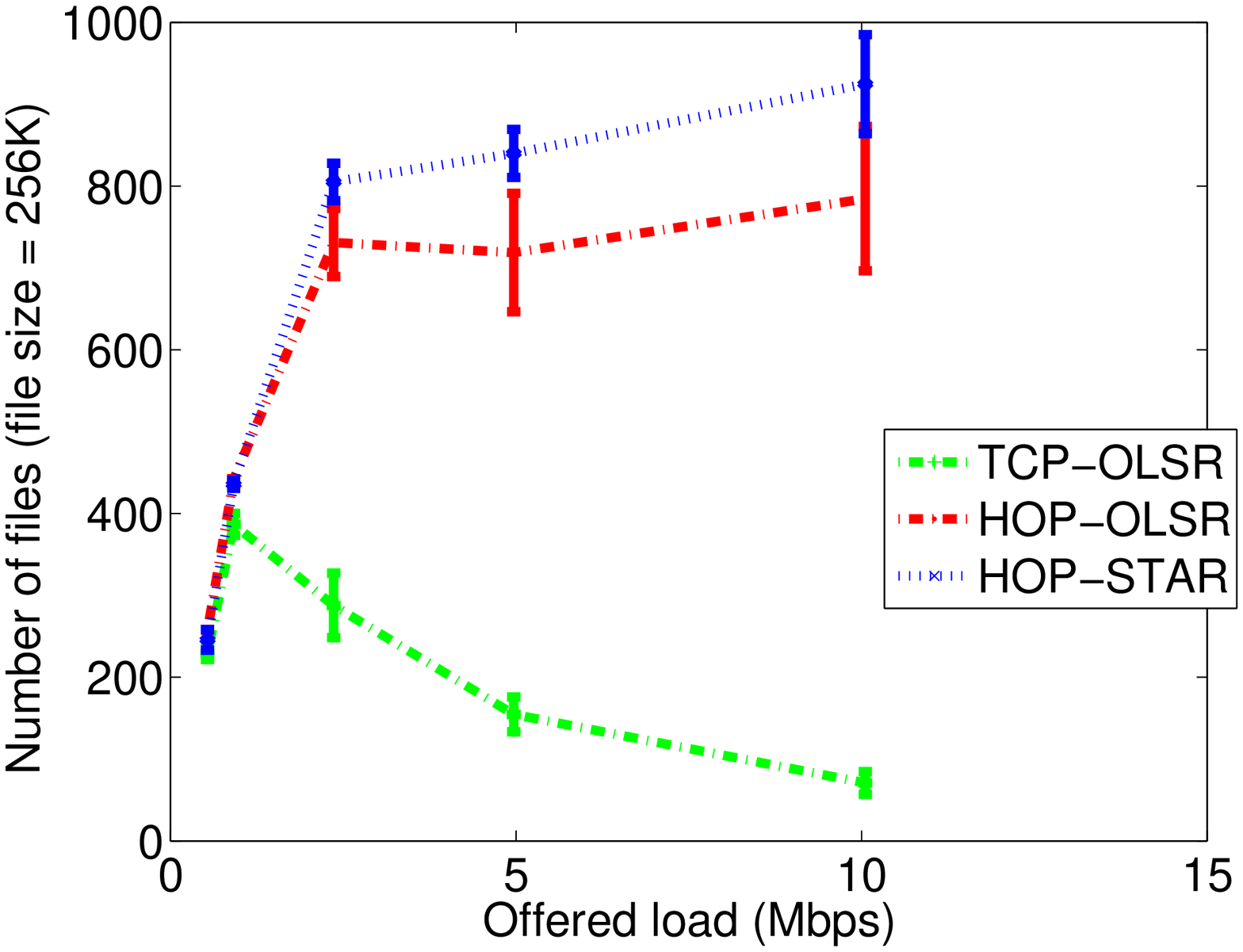}
\label{threshold_delay}
}
\subfigure[Throughout vs. Offered load]{
 \includegraphics[width=0.3\textwidth]{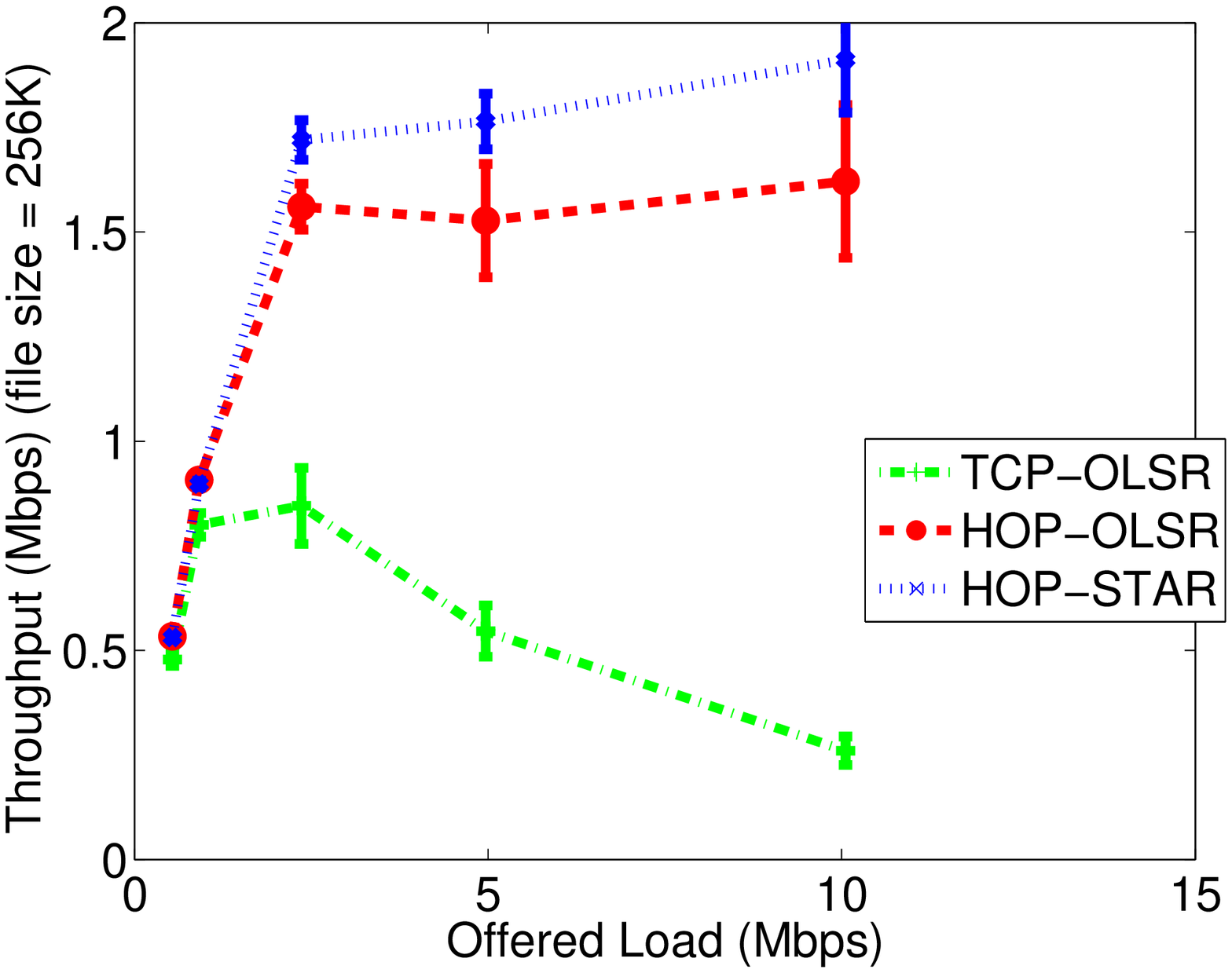}
\label{throughput_load}
}
\caption{Results of Simulation 2a: Static Mesh Network with Exponential traffic model}
\label{sim_mesh_static}
\end{figure*}

\paragraph{Simulation 2: Static Multi-hop Mesh Network}
We simulated a static wireless mesh network with 25 nodes randomly placed in a 500x500m grid. The connectivity graph of a sample network is shown in Figure~\ref{scenario}. Nodes in the network generated traffic at an exponentially distributed rate and the mean inter-arrival rate of files was changed to simulate different  offered loads. Figure~\ref{cdf} shows the CDF of file delays at 10Mbps offered load. ``HOP-STAR'' delivers 932 files which is 17.8$\%$ more compared to ``HOP-OLSR'' and $6.6\times$ compared to ``TCP-OLSR''. Figure~\ref{threshold_delay} shows that at 10Mbps offered load, 924 files in ``HOP-STAR'' are delivered in less than 4 minutes while in ``HOP-OLSR'' and ``TCP-OLSR'' the number is far less (784 and 70 files respectively). Similarly as shown in Figure~\ref{throughput_load}, the overall network throughput achieved in ``HOP-STAR'' is 18\% higher than ``HOP-OLSR'' and 2.25 times more than ``TCP-OLSR'' at 10Mbps offered load. These results show that the storage routing concept helps improve the network performance even when the wireless channel conditions are good but the load in the network is high.

Much of the above performance benefit (as well as in other simulated scenarios) was seen as a result of the preference to store rather than forwarding along paths that are slower than usual. This feature reduces the duration of interference by choosing links that have higher transmission rate and hence are better for spatial reuse if RTS and CTS are not used in the link layer. STAR also prefers storage when when the space in downstream routers is low. This feature acts as a hop-by-hop flow control that prevents data loss due to buffer overflow on downstream routers in high traffic conditions. In comparison the OLSR routing protocol does not have the concept of storage when the path is slower than average or when the downstream routers have lower storage space. In case of ``TCP-OLSR'' where there is an end-to-end flow control implemented by TCP, and congestion control is used to throttle the flow when the network cannot handle the incoming rate. However, as seen in results, the TCP congestion control is too aggressive. The TCP slow start phase completely throttles the rate at which the flow enters the network. This leads to lower throughput as the offered load increases.  
\begin{figure*}[ht]
\centering
\subfigure[CDF of file delays: mean inter-arrival time of files is 1 seconds (offered load = 14Mbps)]{
\includegraphics[width=0.3\textwidth]{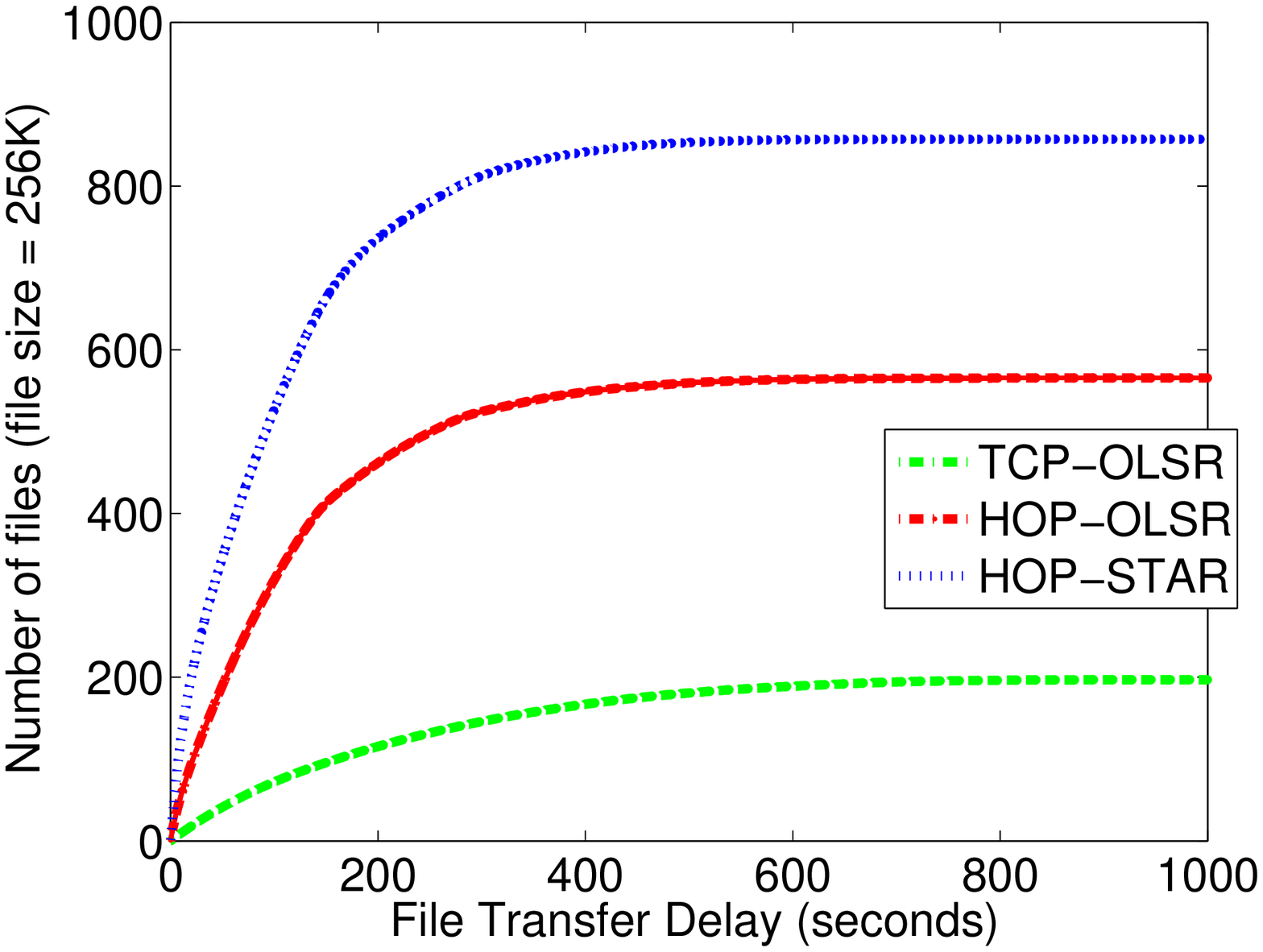}
\label{cdf_b}
}
\subfigure[Number of files delivered with less than 4 minutes delay]{
\includegraphics[width=0.3\textwidth]{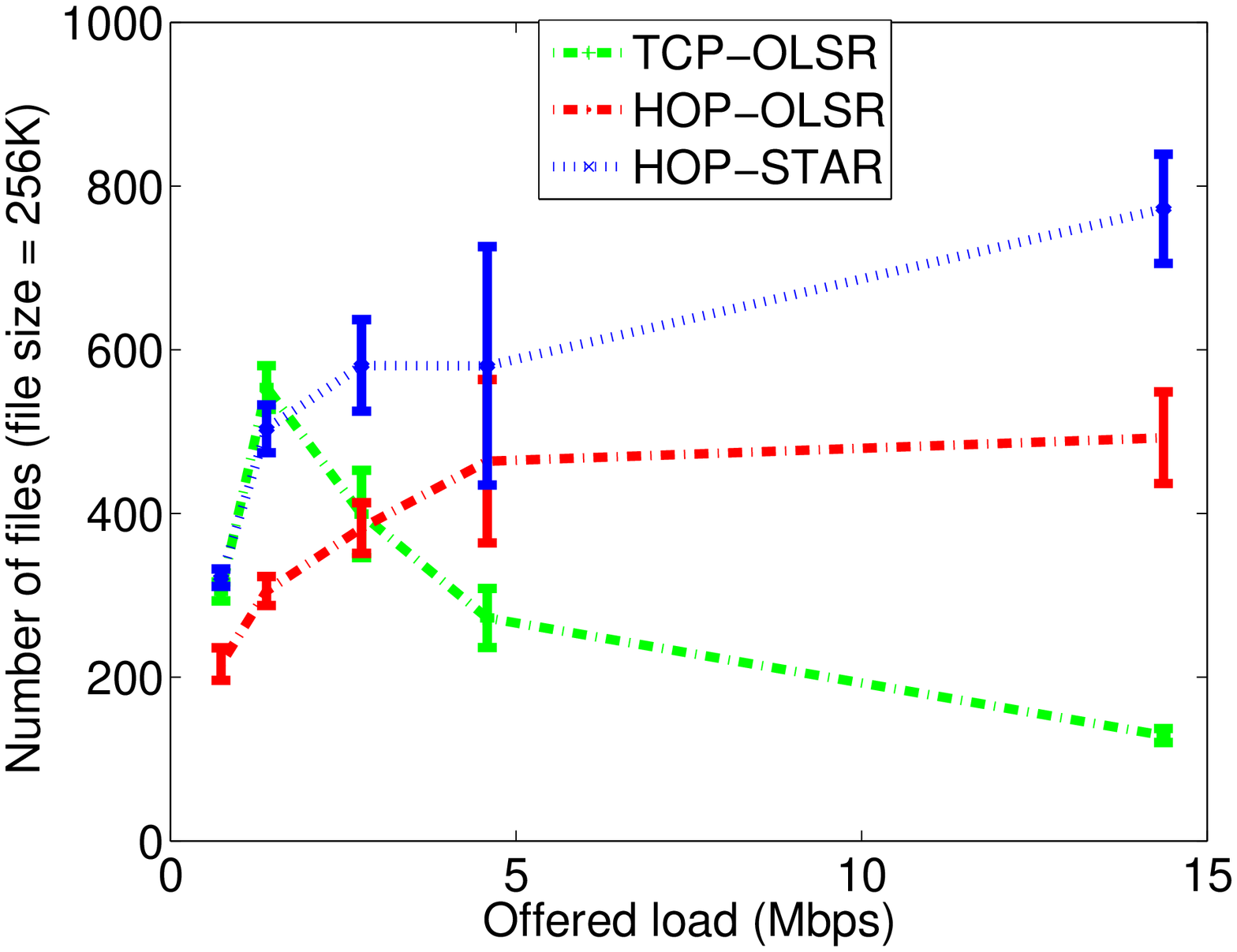}
\label{threshold_delay_b}
}
\subfigure[Throughout vs. Offered load]{
\includegraphics[width=0.3\textwidth]{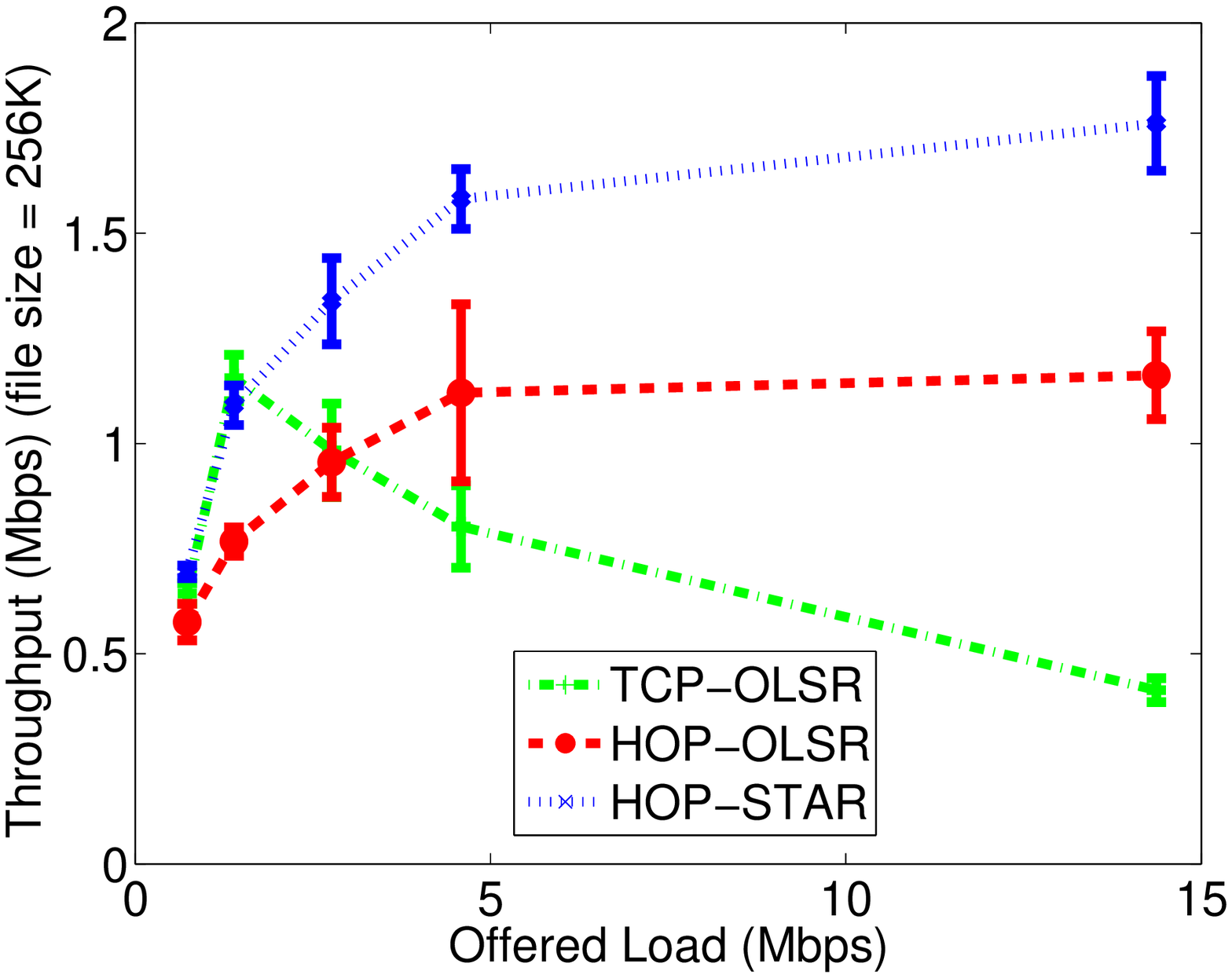}
\label{throughput_load_b}
}
\caption{Results of Simulation 2b: Static Mesh Network with Bursty traffic model with burst durations of 100 seconds and interval between two bursts is 200 seconds}
\label{expt_5_bursty_model}
\end{figure*}
We repeated this simulation under a bursty traffic model. In the same 25 node scenarios as above, we changed the traffic produced by nodes so that they generate traffic by sending files at exponentially distributed inter-arrival rates in 100 second bursts followed by 200 second periods when no new traffic is produced. This process starts randomly at each node and repeats throughout the simulation period. This model more closely represents a web-browsing traffic where users make requests in between pauses. Results (Figure~\ref{expt_5_bursty_model}) show that ``HOP-STAR'' performs better than ``HOP-OLSR'' and ``TCP-OLSR''. The flow control in STAR acts as a traffic regulator by disallowing a large burst of traffic from entering the network at once and hence improves the network performance. TCPs flow control also works better in this scenario by taking advantage of the time periods when no new traffic is generated during the interval between consecutive bursts.

\begin{table}
\centering
\caption{Simulation 3 - Results in the DTN and Sparse MANET scenario:}
\begin{tabular}{|p{4cm}|c|c|c|c|c|c|}
\hline
\multicolumn{1}{|p{4cm}|}{\textbf{Parameters}} & \multicolumn{2}{|c|}{\textbf{OLSR}} &  \multicolumn{2}{|c|}{\textbf{STAR}} \\
\hline
Mean inter arrival time (s) & 10 & 50 & 10 & 50 \\
\hline
File deliver fraction  & 72.34 & 66.67 & 89.66 & 79.17\\
\hline
Average file delay (s) & 100.65 & 109.65 & 92.28 & 38.66 \\
\hline
File streaming throughput(Mbps) & 1.09 & 1.39 & 1.7 & 1.59 \\
\hline
\end{tabular}
\label{table:levy_results}
\end{table}

\begin{figure}[ht]
\centering
\includegraphics[width=0.5\textwidth]{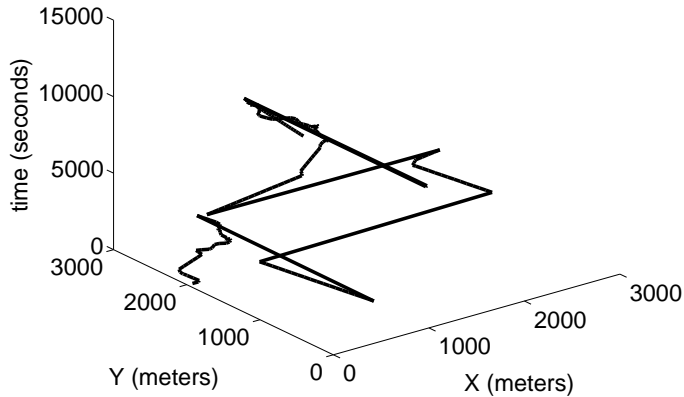}
\caption{Simulation 3 (DTN scenario): Truncated Levy Walk example. Power law slope of flight length = 1 and pause time = 10 seconds}\label{levy_scenario}
\end{figure}
\paragraph{Simulation 3: DTN scenario}
We simulated a 25 node network in which mobile nodes move within a 2500x2500m area. Nodes follow the Truncated Levy Walk (TLW)~\cite{TLW08} mobility model(Figure~\ref{levy_scenario}) which represents pedestrian mobility on a campus. In the absence of infrastructure nodes, this network represents a sparse MANET or a DTN  scenario where end-to-end paths between pair of nodes is not always available. Results  (Table~\ref{table:levy_results}) show that ``HOP-STAR'' can deliver $\approx 80-90\%$ of the files even in these challenged conditions while ``HOP-OLSR'' falls short in file delivery fraction, delay as well as file streaming throughput. This observation shows that  both OLSR and STAR can function in temporarily disconnected scenarios. However, the performance is better when STAR is used.  Compared to OLSR,  STAR can handle occasional outages in sparse networks when nodes may wander away from communication range of the rest of the network leading to temporary network partitions. In our implementation, both OLSR and STAR store data at intermediate nodes when there is no route available and until the mobile nodes move around and reconnect. However, STAR gains an edge over OLSR due to the additional constraint on not using routes when they are worse than average. Thus when a route is found, STAR prefers a route which is more stable or has better link rate compared to the average. This choice results in higher file streaming throughput. Thus there are fewer retransmissions and hence lower interference when weaker than average links are avoided which also means better spacial reuse leading to higher file delivery ratio. These results show that STAR works well in challenged network conditions such as sparse MANET and DTN.
\begin{figure*}[ht]
\centering
\subfigure[ CDF of file transfer delays with mean inter-arrival file transfer rates of 5 seconds (Offered Load = 10Mbps)]{
\label{cdf_mobile}
\includegraphics[width=0.3\textwidth]{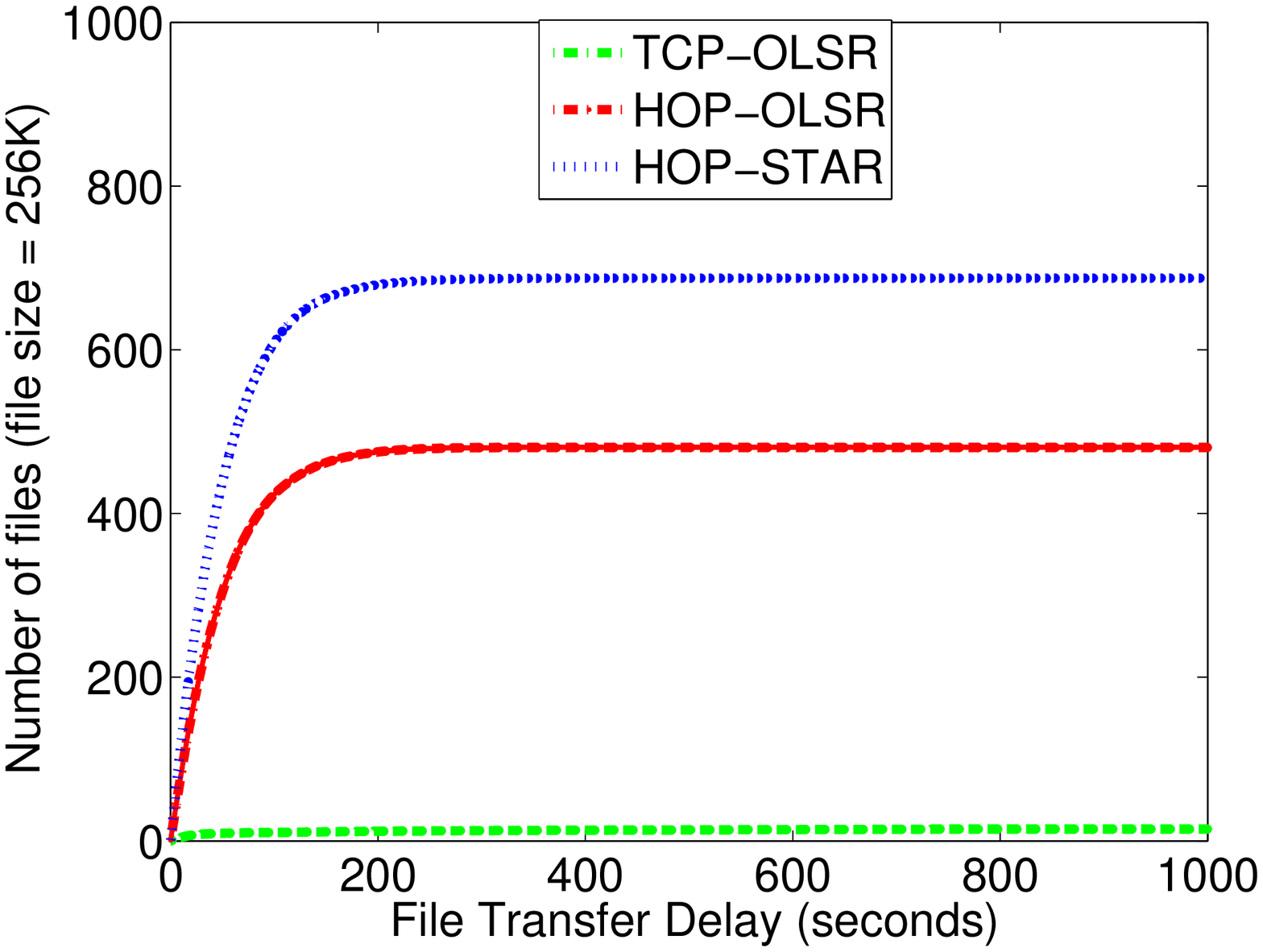}
}
\subfigure[Number of files delivered with less than 4 minutes delay ]{
\label{thresh_mobile}
 \includegraphics[width=0.3\textwidth]{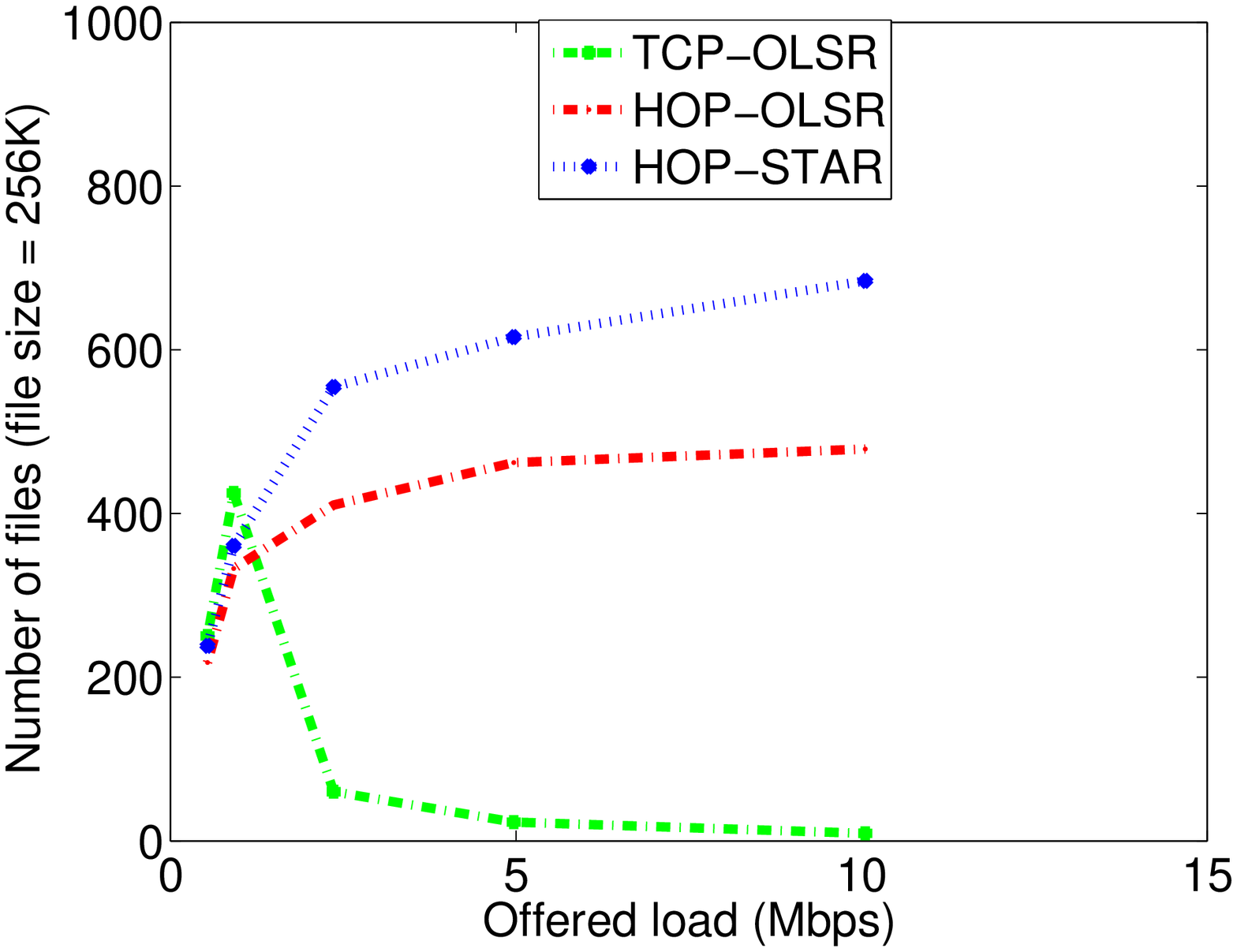}
}
\subfigure[Throughput vs Offered load]{
\label{load_mobile}
\includegraphics[width=0.3\textwidth]{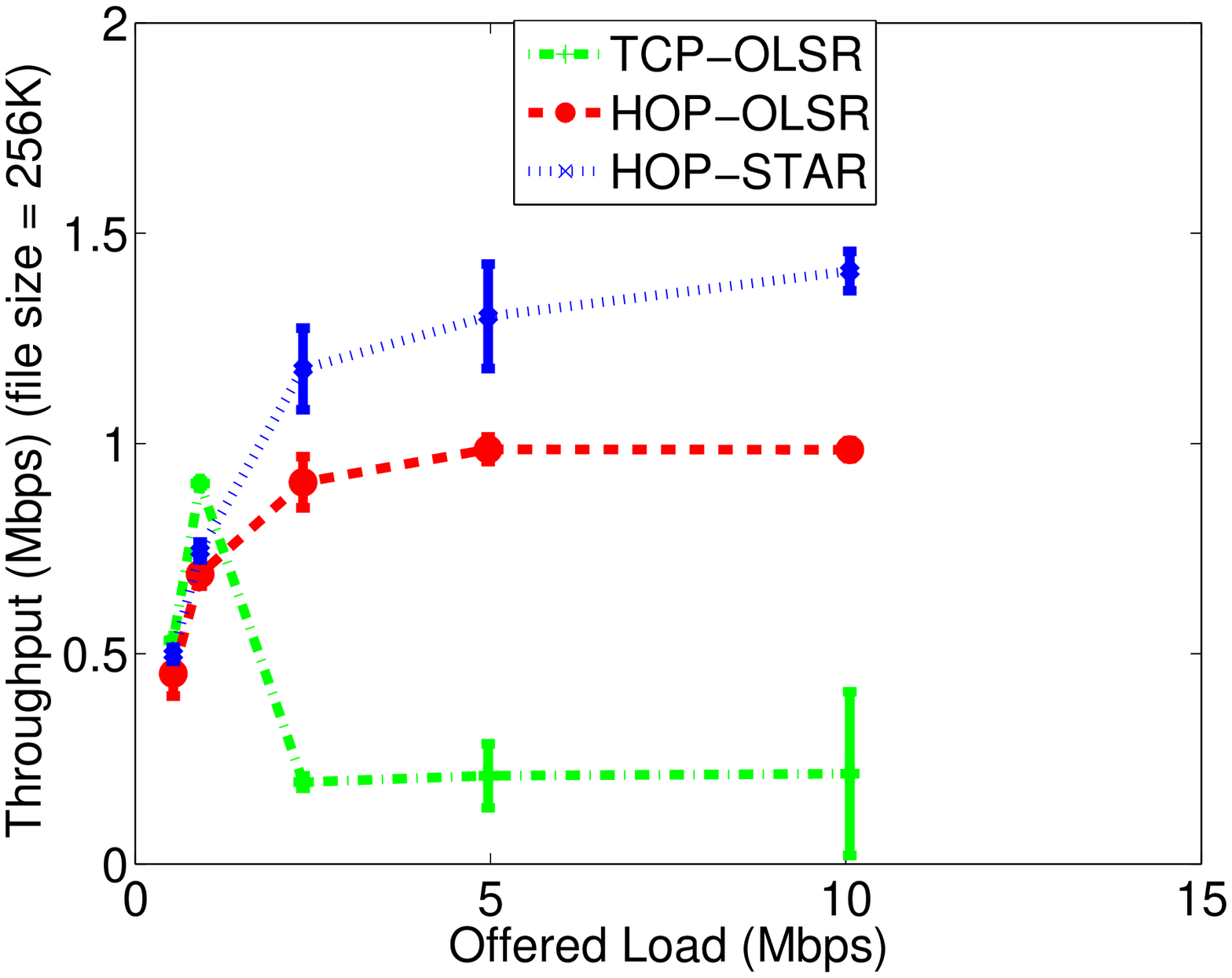}
}
\caption{Results of Simulation 4a: Manhattan Mobility Model with exponentially distributed file tranfer rates}\label{expt_4_man_mob}
\end{figure*} 
\paragraph{Simulation 4 - Manhattan Mobility Scenario: }
We simulate a mobile scenario with 25 nodes within a 500x500m grid. The grid consists of horizontal and vertical streets and mobile nodes move along the streets. At street intersections nodes may turn right, left or go straight with equal probabilities. The minimum and maximum node speed in this network was set to 5 and 10m/s respectively. This is known as the Manhattan mobility model~\cite{man_mob} and it emulates vehicular mobility in an urban environment. Results (Figure~\ref{expt_4_man_mob}) show that ``HOP-STAR'' delivers between 250 to 300 more files with less than 4 minute delay compared to ``HOP-OLSR''. ``HOP-STAR'' also achieve higher throughput compared to ``TCP-OLSR''.
\setcounter{paragraph}{0}  
\subsection{Proof-of-Concept Testbed Implementation} We have evaluated the ``HOP-STAR'' protocol stack and compared with ``HOP-OLSR'' and ``TCP-OLSR'' on the ORBIT~\cite{orbit05} wireless testbed to provide a proof-of-concept in actual implementation. We  used the HOP implementations provided by Li et al~\cite{hop_umass} and the OLSR implementation with link cost extensions by Tomp~\cite{ett_olsr} respectively as the baseline for the implementation. The expected transmission time (ETT) metric~\cite{ett_metric} is used to compute path cost in both OLSR and STAR. In STAR, the most recently learned ETT is used as the short term cost and a moving average of ETT is maintained as the long term cost. STAR also uses storage information on downstream path to decide whether data should be forwarded or temporarily stored at the current router. As shown in results before, this storage consideration acts as a flow control mechanism that restricts the influx of flow into the network and hence reduces the chance for buffer overflows at downstream routers. In addition, the 2-D forwarding strategy makes opportunistic use of good links. OLSR, however, does not have such mechanism built except that if there is no available route, data is stored at the transport layer at each hop instead of being dropped immediately. We performed experiments in three scenarios and the results are presented in this section.  
\begin{figure}[!t]
\centering
 \subfigure[Topology]{
  \includegraphics[width=0.5\textwidth]{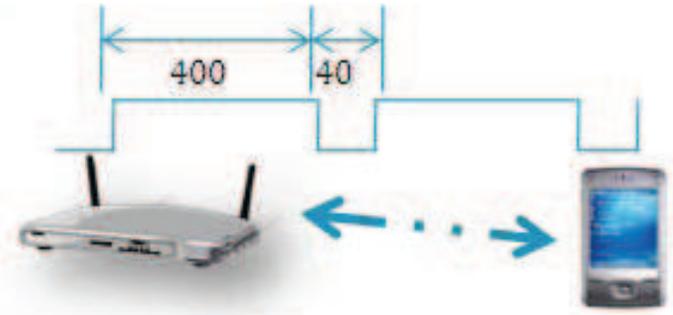}
  \label{fig:orbit_topology_1}
  }
  \subfigure[Result]{
  \includegraphics[width=0.5\textwidth]{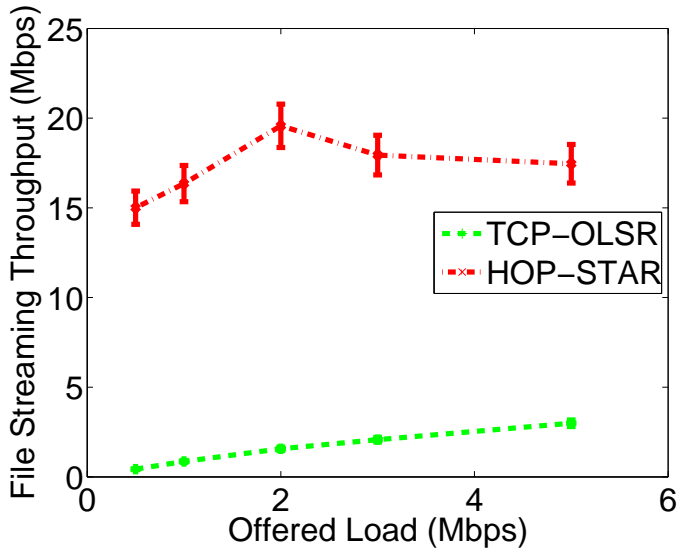}
  \label{fig:orbit_results_1}
  }
  \caption{Testbed Experiment 1: Mobile node and Access Point Scenario}
\end{figure}
\paragraph{Testbed Experiment 1 - Mobile node and Access Point:} We emulate a scenario in which a mobile device connects to an access point (AP) to download content (Figure~\ref{fig:orbit_topology_1}). Mobility directly results in SINR variations and 
leads to variable bit error rate (BER) between the transmitter and the receiver. The transmitter reacts to higher error rates by reducing the transmission rate of the link. Therefore, we emulate mobility by periodically switching the transmission rates of the wireless driver between 1Mbps to 11Mbps as shown in Figure~\ref{fig:orbit_topology_1}. We experiment with two protocol stacks: $\{$HOP,STAR,802.11g$\}$ 
and $\{$TCP,OLSR,802.11g$\}$. Traffic is generated by sending data from the AP to the mobile at 5 second intervals. Size of application layer data is varied from 325-1024 KBytes to increase the offered load in the network. Results (Figure~\ref{fig:orbit_results_1}) show that STAR is able to achieve 20Mbps file streaming throughput which is only 5Mbps short of the theoretical maximum for a 54Mbps data rate with 1KByte  limit on the link layer frame size (Figure 2 in~\cite{Xiao_theretical_802.11} shows that 25Mbps is the theoretical maximum). In comparison, TCP  provides a mere 3Mbps throughput. Note that the streaming throughput in TCP is equivalent to the network throughput as there is no concept of storage in TCP. This result may be explained as follows: TCP reacts to lower link rates by dropping the congestion window to the minimum by  entering the slow start phase. However, the HOP and STAR combination continues to send at the full rate when the link rate is high and temporarily suspends the transmissions and stores the data when the link rate reduces. The switch back to transmission at high speed is much faster than the time taken by the TCP slow start process since it directly uses the network layer feedback to adjust the transmission rate. Therefore, overall HOP and STAR achieve a higher streaming throughput compared to TCP.  
\begin{figure}[!t]
\centering
 \subfigure[Topology]{
 \label{fig:orbit_topology_2}
 \includegraphics[width=0.5\textwidth]{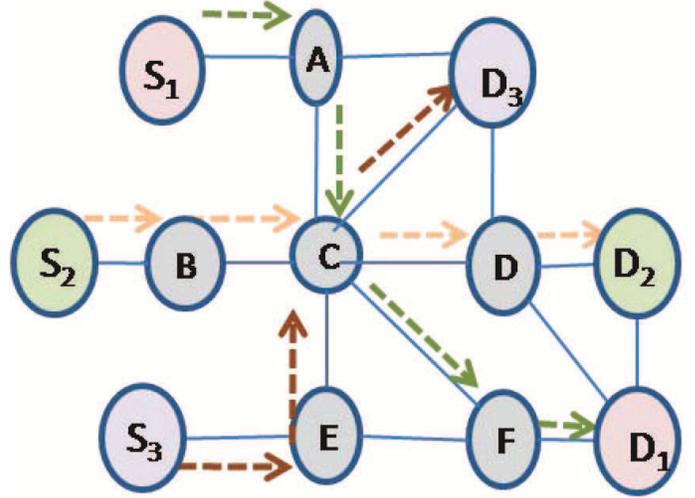}
 }
 \subfigure[Results]{
  \label{fig:orbit_results_2}
  \includegraphics[width=0.5\textwidth]{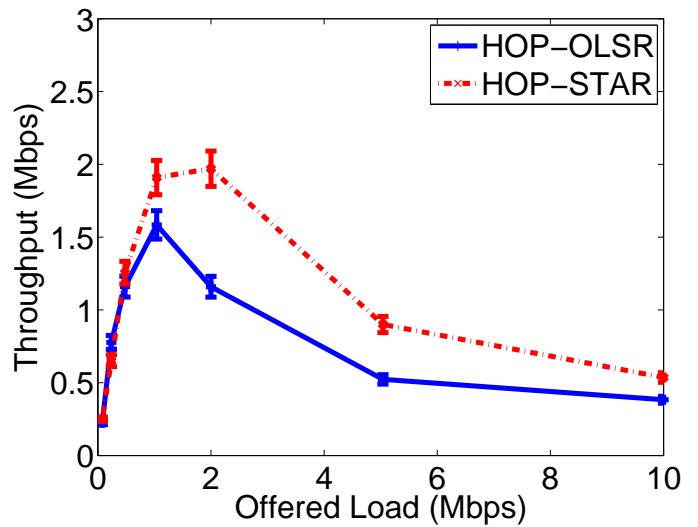}
  }
  \caption{Testbed Experiment 2: Wireless Mesh Network}
\end{figure}
\paragraph{Testbed Experiment 2 - Multihop Mesh Network: } In the second experiment, we show how the storage based forwarding decision in STAR helps to achieve better performance in a multihop mesh network. We create a 12 node multihop topology (Figure~\ref{fig:orbit_topology_2}) in which three sources $S_1$,$S_2$ and $S_3$ transmit files of size 500KB each to corresponding destinations $D_1$,$D_2$ and $D_3$. Each source generates new file transfers at an exponentially 
distributed rate with mean $\lambda$ files per second. We increase $\lambda$ at each node to increase the offered load 
in the network. Two protocol stacks $\{$HOP,OLSR,802.11g$\}$ and $\{$HOP,STAR,802.11g$\}$ were 
compared. Figure~\ref{fig:orbit_topology_2} shows that all files are transferred 
along the three paths that pass through node C. As the network load increases, the storage space 
at node C starts filling up. When upstream nodes in STAR detect the low storage levels along the 
route, they choose to store instead of forwarding. However, since OLSR does not use storage information to make forwarding decisions, it continues to push data along the same route which may result in buffer overflow at node C when offered load is sufficiently high. Therefore, due to the storage metric used in STAR, a better throughput performance is achieved compared to OLSR (Figure~\ref{fig:orbit_results_2}).
\begin{figure}[!t]
\centering
 \subfigure[Topology]{
  \includegraphics[width=0.5\textwidth]{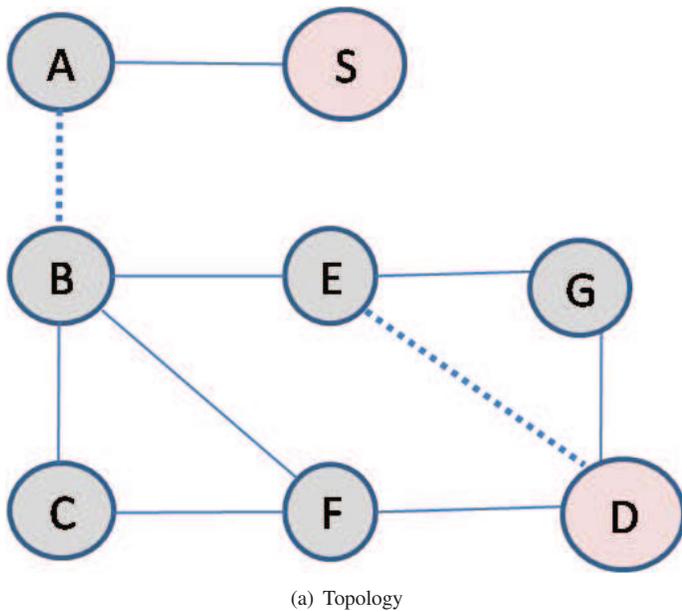}
  \label{fig:orbit_topology_3}
  }
  \subfigure[Result]{
  \includegraphics[width=0.5\textwidth]{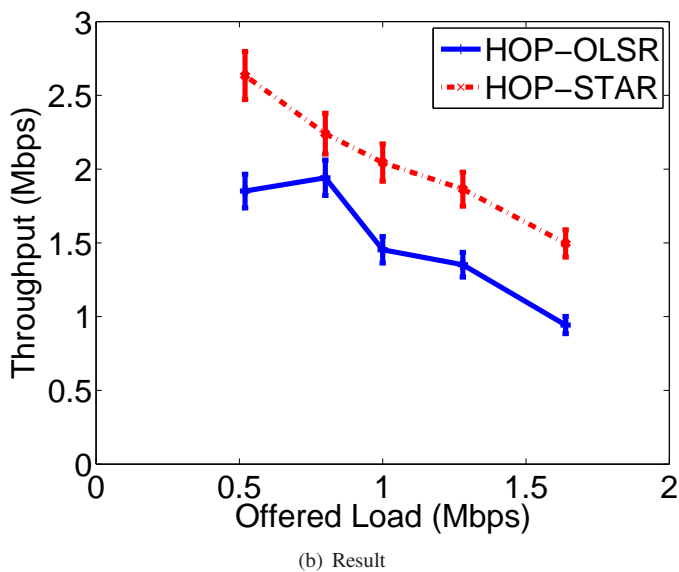}
  \label{fig:orbit_results_3}
  }
  \caption{Testbed Experiment 3 - Wireless Mesh Network with periodically disconnected links: }
\end{figure}
\paragraph{Testbed Experiment 3: Multihop Mesh with Periodic Disconnections}
This experiment validates our claim that STAR can work well even in DTN scenario when the end to end paths between the source and destination do not always exist. We create a scenario shown in Figure~\ref{fig:orbit_topology_3} where the links A-B and E-D are periodically turned off and on for 40 and 200 seconds respectively. The off states of these links are phase shifted by 150 seconds so that the two links are never ``off'' at the same time during the experiments. Files are sent at a constant rate of 1 file in 5 seconds from source S to destination D. STAR works as follows in this network. When the link A-B is off, the long term cost for the route is finite while the short term cost is infinite. A comparison of the two costs indicates that a route to D existed in the past. STAR uses this information to temporarily store the file while waiting for the link to reappear. Similarly, when the link E-D is off, STAR is able to route data to D through the alternate path that goes through G. On the other hand, when the link E-D is on, STAR opportunistically selects the lower cost path to reach D. As seen in Figure~\ref{fig:orbit_results_3}, these choices validate that STAR can function well in challenged DTN scenarios.  TCP does not function at all in this scenario therefore we have not included the results for ``TCP-OLSR''.  We show that OLSR functions in this scenario if a hop-by-hop transport layer is used because the transport layer also stores data if there is no route available. However, OLSR is unable to switch routes when the storage at downstream routers is low leading to lower throughput compared to STAR. 

\section{Related Work}\label{sec:related}
\setcounter{paragraph}{0}
In this section we present related work in three areas. First we describe routing protocols proposed in future Internet architectures and explain how STAR fits in. Second, we present the difference between DTN routing and STAR. Third, we compare STAR with prior work in a storage routing protocol~\cite{storage_dtn} and a history based routing~\cite{hidra} protocol.
\paragraph{Future Internet Architectures}
\eat{A common theme that ties future Internet architecture proposals is the use of names rather than  location. This choice automatically supports mobility, security and privacy. In this section, we present three architectures: Named Data Networks,  eXpressive Internet Architecture and MobilityFirst. We also present Content Centric Networking and the Forwarding directive, Association, Rendezvous Architecture (FARA)~\cite{fara} architectures where the basic principles of  NDN and XIA respectively can be found. We show the position of STAR as a network routing protocol for each architecture and discuss any adaptation that might be needed.

  The \emph{Content Centric Network (CCN)} is the parent idea behind the \emph{Named Data Network (NDN)}~\cite{ndn} architecture. CCN was proposed by Jacobson et al in~\cite{ccnx} to replace the conversation based content retrieval model of the Internet to a content caching and content multicast based model. In CCN, a host $A$ requests content $B$ by sending an interest packet containing the names $B$ and $A$ through all network interfaces. A router that receives this interest packet, looks into its cache to find the content $B$. If the content is found, the router simply sends the content to the interface from which the interest arrived. If however, the data is not available in the cache, the router first updates its Pending Interest table(PIT) and then forward the interest along all other interfaces. The PIT is updated as follows: If there is an entry for $B$ in the PIT, the router checks if the interface through which this interest arrived is already present in this entry, otherwise the entry is updated to include that interface. If their is no entry for $B$ in the PIT, a new entry is added and the interface number is noted as the source of the interest for $B$ in the entry.  When the router receives content named $B$, it caches the content and looks up the PIT to find an entry for $B$. If such an entry is found, the router forwards $B$ along all interfaces $i$ which are noted as the sources of the content. This router may now serve all future interests for $B$ from its own
cache.  Following the same interest and data based ``narrow-waist'', NDN implements naming schemes, security of content at each router and support for mobility. Routers in NDN need several levels of storage space for caching and maintaining tables, which our reference storage router design can easily accommodate.} 
The Named Data Network architecture proposes an outline of a link state routing scheme (NLSR) that maintains multiple paths to all known destinations and the routes are ranked in any suitable manner to indicate preferences. Best paths are computed using the Dijkstra's algorithm over the entire sink tree rooted at node $A$. The Dijkstra's algorithm is repeated over new the sink trees created by removing one neighbor of $A$ at a time from the original sink tree. This process may generate a set of alternate paths to some nodes. An upper bound on the number of such computations is used to reduce the complexity at high degree nodes. This is only a suggested multi-path algorithm while other schemes might be explored in the future.  STAR could be easily used in NDN by simply replacing the messaging scheme (``hello'' and ``TC'' messages) to use the NDN narrow waist i.e., interest for link state updates followed by the corresponding responses. There already exists a multi-path computation scheme in STAR and the preferences for route selection is modeled as well. Therefore, other than the changes to control messaging scheme, the STAR concept is  applicable to the NDN architecture. Aside from ``fitting in'' as an NDN routing protocol, STAR adds the support for mobility more natively than is currently suggested for NDN. Grassi et al~\cite{grassi2013vehicular} suggest stripping out all NDN features e.g., forwarding information base and pending interest table policy and resort to epidemic routing~\cite{vahdatand2011epidemic} like route computation to use NDN in vehicular networks. In comparison, as shown in results in Section~\ref{sec:sim}, STAR works well in both disconnected scenarios as well as vehicular mobility scenarios with the same routing protocol mechanism as the well connected static mesh networks. Similarly, to support ad-hoc networking in NDN, a Listen First Broadcast Later (LFBL) approach is suggested~\cite{meisel2010ad}. This approach is reminiscent of reactive routing in ad-hoc networks with some enhancements to reduce broadcast floods. There are several studies that compare proactive and   reactive routing approaches in ad-hoc networks~\cite{mbarushimana2007comparative, lee2004proactive}. Results are very much dependent on the scenario, application and mobility models. Therefore, STAR adapted to an NDN based ad-hoc network might work atleast as well as LFBL.

 The \emph{Forwarding directive, Association, Rendezvous Architecture (FARA)}~\cite{fara} was proposed as an addressing and naming scheme for the future Internet. In FARA, hosts are identified by an association ID (AID) instead of IP addresses. In addition, hosts are not individual machines, rather an entity in the form of a process or thread running anywhere in the network. Entities may move around in the network and may even migrate from one physical machine to another. Routing is achieved by constructing forwarding directives(FD) that lead to the destination entities. FDs are updated as the entities move around in the network. When a router receives a data packet, it checks the FD of the corresponding destination entity to make the forwarding decision. FARA does not address protocol details like missing FDs, delay in FD update message and variation in routing costs in the network. STAR can provide this service by temporary storage and re-routing of data as necessary. The \emph{eXpressive Internet Architecture (XIA)} builds on similar ideas to construct a network where service, content, host or any other principle (entity) can be expressed in a network packet. The core research theme is in naming, addressing, privacy, security, incentives and software implementation. The network routing protocol and algorithm have not been locked down yet. The only requirement for routing in XIA is to locate the entities in the network. STAR can be plugged in the XIA architecture as the routing protocol for all scenarios after changing the control packet headers to enable routing to different XIA principles. Separate tables can be maintained for different entities while using the same number and types of control messages for all types of entities. 

The \emph{MobilityFirst architecture} has already adapted STAR as the generalized storage aware routing protocol (GSTAR)~\cite{nelson2011gstar}. Extension of STAR in multi-homed networks is a work in progress~\cite{zhangsupport}. The main ideas of GSTAR continue to be multipath computation, two-dimensional routing cost metric, and storage aware forwarding decisions. These ideas are extended by using a minimum delay as the route cost metric~\cite{nelson2011gstar} and simultaneous use of multiple paths to improve content delivery to multi-homed mobiles~\cite{zhangsupport}. 
\paragraph{Disruption/Delay Tolerant Networking (DTN)}
The goal of DTN routing~\cite{Mirco:05,Lindgren:03} is to successfully deliver messages to a destination which may not be always connected. Both DTN and STAR routers are designed with built-in storage. In DTN replication, flooding and contact probability are three major approaches for routing while STAR uses a hop distance routing metric and a variable cost metric based on packet transmission times. The variable metric is able to capture various states of disconnections and hence is capable of functioning under deep disconnections. Although not designed explicitly for DTN, STAR with the two dimensional routing metric and storage based forwarding destination functions well in challenged scenarios such as DTN and sparse MANET as well as in well connected ones.

The Licklider transport protocol (LTP)\cite{ramadas2008licklider} was designed to support reliable transport along links that have large round trip delays or long term disconnections. The DTN bundle protocol is based on the LTP protocol which makes LTP the precursor of DTN. LTP is best deployed right above the link layer where it is used to provide minimal transport functionality of end-to-end reliability over challenged links. LTP can also be implemented above UDP but RFC5326~\cite{ramadas2008licklider}, that describes the specifications in details, advises against using LTP above UDP in anything other than {\it during software development and in private networks}. LTP needs link layer information in order to make store or forward decisions.  However, LTP does not participate in building routes or maintaining link quality information. We differentiate our work from LTP as follows: First LTP is a transport protocol while we have proposed a network routing protocol. Our protocol, STAR, provides a best effort, rather than reliable delivery service to the transport layer. Second, LTP does not provide any flow or congestion control while the forwarding algorithm in STAR performs network layer backpressure based flow and congestion control by storing data when the network is slow or when  the storage at downstream routers is low. Third, unlike LTP, STAR  performs route discovery and table maintenance to log link quality variations. Finally, LTP is not recommended for use in scenarios other than those that have long haul links while we show that STAR is capable of functioning in a wide range of scenarios.

\paragraph{Storage and History based routing}
Storage Routing (SR) has also been proposed for DTN networks~\cite{storage_dtn} but the focus is on managing the storage space on routers when there are too many undelivered bundles at a router. SR suggests a protocol for moving old messages to neighboring routers to make room for new arrivals while the older message bundles wait to encounter a suitable node that can take custody for further forwarding. This concept is quite different from the way storage is used in STAR. STAR uses storage for disconnected users as well as when the path to the destination is slower than usual. Storage information on downstream routers is also used to restrict data forwarding which leads to  a network layer flow control that prevents buffer overflows at downstream routers. History directed routing algorithm(HIDRA)~\cite{hidra} has been suggested for multi-media traffic in the IP network. In HIDRA the general range of bandwidth between source and destination pairs are observed during data transfer sessions. These measurements are processed offline and used to select the best routes for future traffic between the hosts. The historical information is used to predict the load between a source destination pair and to assign the flow to the best route in the network that can accomodate the flow. In this respect, HUDRA is an application of the knapsack or bin-packing problem that tries to allocate and fit incoming flows in the most optimal path.  Unlike HIDRA, STAR is a distributed protocol that constructs the historical information about links and routes even when there has not been any prior data traffic along the route. The link cost metrics are instantaneous and moving average of transmission times. A comparison of current and average metrics are used to decide whether the route should be used immediately or data forwarding should be delayed to wait for the link cost to reduce. The objective in STAR is to enable seamless routing across various network conditions which is different from HIDRA that attempts to improve network throughput in IP networks by allocating incoming flows to most suitable routes. 
\section{Conclusions and Future Work}\label{sec:conclusion}
We have presented STAR: a storage aware routing protocol that considers historical and current routing costs to make store or forward routing decision. Through extensive simulation under different network and mobility models and experimental validation on ORBIT, we show that STAR is a suitable routing protocol for generalized DTN scenarios and functions well even under challenged DTN networks. Results present sufficient evidence that the 2-dimensional routing metric and storage concept can develop as a unified routing protocol that seamlessly inter-connects several types of wireless and mobile networks. We have briefly mentioned the applicability of the principle ideas in STAR i.e., two dimensional routing metric and storage considerations in the NDN and XIA future Internet architectures. We see the implementation of STAR as an NDN routing protocol and performance evaluation in NDN wired, vehicular and ad-hoc networks as the immediate future direction. The STAR idea in newer scenarios such as multi-homing is already a work in progress~\cite{zhangsupport}.  
\eat{
We see two immediate future directions to extend our work on storage aware routing. First, STAR provides a framework for making intelligent forwarding decisions and we have presented a simple heuristics based strategy as an instance. More sophisticated strategies that use reinforcement learning or pattern recognition may be implemented using the 2-D routing cost framework. Second, further work is required to evaluate the performance of STAR in more general wired and wireless network topologies. As the next step we are working on further validation of STAR for a hybrid Wi-Fi + WiMAX scenario using experimental capabilities of GENI~\cite{GENI} testbeds.}
\eat{\section{Acknowledgements}
We would like to acknowledge Snehapreethi Gopinath for her contributions to the project
} 
\section{Bibliography}
\bibliographystyle{elsarticle-num}
\bibliography{star_conext,CISE_bib,ndn}
\eat{
\begin{IEEEbiographynophoto}{S. Jain}
Shweta Jain is an Assistant Professor in the Mathematics and Computer Science department at York College of CUNY and a Doctoral faculty of Computer Science in The Graduate Center of CUNY. She received her BE degree in Electronics and Telecommunication Engineering from Bengal engineering college in 2001, and MS and Ph.D. in Computer Science from Stony Brook University in 2005 and 2007. Her research interests are in mobile and wireless communication and networks. \end{IEEEbiographynophoto}
\begin{IEEEbiographynophoto}{S. Gopinath}
Snehapreethi Gopinath received her Masters in Science in Electrical and Computer Engineering from Rutgers University in 2010. Her master's thesis was on Performance evaluation of the cache and forward routing protocol in multihop wireless subnetworks. She is currently a software engineer at a leading financial company in New York City.  
\end{IEEEbiographynophoto}
\begin{IEEEbiographynophoto}{D. Raychaudhuri}
Dipankar Raychaudhuri is Distinguished Professor, Electrical \& Computer Engineering and Director, WINLAB (Wireless Information Network Lab) at Rutgers University. As WINLAB's Director, he is responsible for an internationally recognized industry-university research center specializing in wireless technology. He is also PI for several large U.S. National Science Foundation funded projects including the "ORBIT" wireless testbed and the “MobilityFirst” future Internet architecture. Dr. Raychaudhuri has previously held corporate R\&D positions including: Chief Scientist, Iospan Wireless (2000-01), AGM \& Dept Head, NEC Laboratories (1993-99) and Head, Broadband Communications, Sarnoff Corp (1990-92).  He obtained the B.Tech (Hons) from IIT Kharagpur in 1976 and the M.S. and Ph.D degrees from SUNY, Stony Brook in 1978 and 79. \end{IEEEbiographynophoto}
}
\end{document}